\documentclass[%
 reprint,
 onecolumn,
 superscriptaddress,
 amsmath,amssymb,
 aps,
]{revtex4-2}

\usepackage{chemformula} 
\usepackage{siunitx}
\usepackage{xr}
\usepackage{subfiles}

\usepackage{graphicx}
\usepackage{dcolumn}
\usepackage{bm}
\usepackage{tabularx}
\usepackage{xcolor}
\usepackage{braket}

\makeatletter
\g@addto@macro\@floatboxreset\centering
\makeatother

\makeatletter
\newcommand*{\addFileDependency}[1]{
  \typeout{(#1)}
  \@addtofilelist{#1}
  \IfFileExists{#1}{}{\typeout{No file #1.}}
}
\makeatother

\begin{document}

\title{Electronic and Excitonic Properties of MSi$_{\textrm{2}}$Z$_{\textrm{4}}$ Monolayers}

\author{Tomasz Woźniak}
\email{tomasz.wozniak@pwr.edu.pl}
\affiliation{Department of Semiconductor Materials Engineering, Wrocław University of Science and Technology, 50-370 Wrocław, Poland.}
\affiliation{Department of Physics and Earth Sciences, Jacobs University Bremen, Campus Ring 1, 28759 Bremen, Germany}
\affiliation{Helmholtz-Zentrum Dresden-Rossendorf, Abteilung Ressourcenökologie, Forschungsstelle Leipzig, Permoserstr. 15, 04318 Leipzig, Germany}

\author{Umm-e-hani Asghar}
\affiliation{Department of Physics and Earth Sciences, Jacobs University Bremen, Campus Ring 1, 28759 Bremen, Germany}

\author{Paulo E. Faria Junior}
\affiliation{Institute for Theoretical Physics, University of Regensburg, Universitätsstraße 31, 93040 Regensburg, Germany}

\author{Muhammad S. Ramzan}
\affiliation{Department of Physics and Earth Sciences, Jacobs University Bremen, Campus Ring 1, 28759 Bremen, Germany}
\affiliation{Helmholtz-Zentrum Dresden-Rossendorf, Abteilung Ressourcenökologie, Forschungsstelle Leipzig, Permoserstr. 15, 04318 Leipzig, Germany}
\affiliation{Institut für Physik, Carl von Ossietzky Universität Oldenburg, 26129 Oldenburg}

\author{Agnieszka B. Kuc}
\email{a.kuc@hzdr.de}
\affiliation{Department of Physics and Earth Sciences, Jacobs University Bremen, Campus Ring 1, 28759 Bremen, Germany}
\affiliation{Helmholtz-Zentrum Dresden-Rossendorf, Abteilung Ressourcenökologie, Forschungsstelle Leipzig, Permoserstr. 15, 04318 Leipzig, Germany}


\begin{abstract}

$MA_2Z_4$ monolayers form a new class of hexagonal non-centrosymmetric materials hosting extraordinary spin-valley physics. While only two compounds (MoSi$_2$N$_4$ and WSi$_2$N$_4$) were recently synthesized, theory predicts interesting (opto)electronic properties of a whole new family of such two-dimensional materials. Here, the chemical trends of band gaps and spin-orbit splittings of bands in selected $M$Si$_2Z_4$ ($M$ = Mo, W; $Z$ = N, P, As, Sb) compounds are studied from first-principles. Effective Bethe-Salpeter-equation-based calculations reveal high exciton binding energies. Evolution of excitonic energies under external magnetic field is predicted by providing their effective $g$-factors and diamagnetic coefficients, which can be directly compared to experimental values. In particular, large positive $g$-factors are predicted for excitons involving higher conduction bands. In view of these predictions, $M$Si$_2Z_4$ monolayers yield a new platform to study excitons and are attractive for optoelectronic devices, also in the forms of heterostructures. In addition, a spin-orbit induced bands inversion is observed in the heaviest studied compound, WSi$_2$Sb$_4$, a hallmark of its topological nature.

\end{abstract}

\maketitle

\section{Introduction}

Two dimensional (2D) materials attract significant interest due to their extraordinary properties and numerous potential applications.\cite{Khan2020} The isolation of graphene~\cite{Novoselov2004} and MoS$_\textrm{2}$~\cite{Mak2010} monolayer (1L) from their parental bulk forms stimulated a wide range of experimental and theoretical investigations, starting from (opto)electronic properties, through spin-orbit physics, to creation of van der Waals (vdW) heterostructures and twist-angle dependent properties.

Among 2D vdW materials investigated up to date, 1L transition-metal dichalcogenides (TMDCs) in the 2$H$ polytype (also known as $H^h_h$) are especially appealing for optoelectronic devices, thanks to their direct band gaps in visible and near-infrared range, and large exciton binding energies. Moreover, their symmetry and strong spin-orbit coupling (SOC) implies a coupling between spin and momentum of carriers in the $K$ valleys of hexagonal Brillouin zone (BZ), which is a central property for their spintronics and valleytronics applications.\cite{Mueller2018,Wang2018}
These properties of 1L TMDCs encouraged scientists to look for new families of similar materials.

Recently, Hong \textit{et al.}~\cite{Hong2020a} proposed a new concept of chemical vapour deposition growth of non-layered materials with passivation of their high energy surfaces. This enabled synthesis of centimeter-scale 1Ls of MoSi$_2$N$_4$ and WSi$_2$N$_4$ compounds.
Monolayer MoSi$_2$N$_4$ was experimentally shown to exhibit semiconducting properties with fundamental indirect band gap of 1.94~eV and excitonic A and B transitions of, respectively, 2.21~eV and 2.35~eV, originating from direct interband transitions at $K$ valleys.
Large carrier mobilities (270 and 1200~cm$^2$~V$^{-1}$~s$^{-1}$ for electrons and holes), on-off ratio of 4000, and high mechanical stability (Young’s modulus and breaking strength of 491.4~$\pm$~139.1~GPa and 65.8~$\pm$~18.3~GPa), superior to 1L MXenes and MoS$_2$ analogues, were reported.\cite{Hong2020a}
Furthermore, the samples were found to be robust under ambient conditions with no surface oxidation. 

\begin{figure*}[ht!]
 \includegraphics[width=0.7\columnwidth]{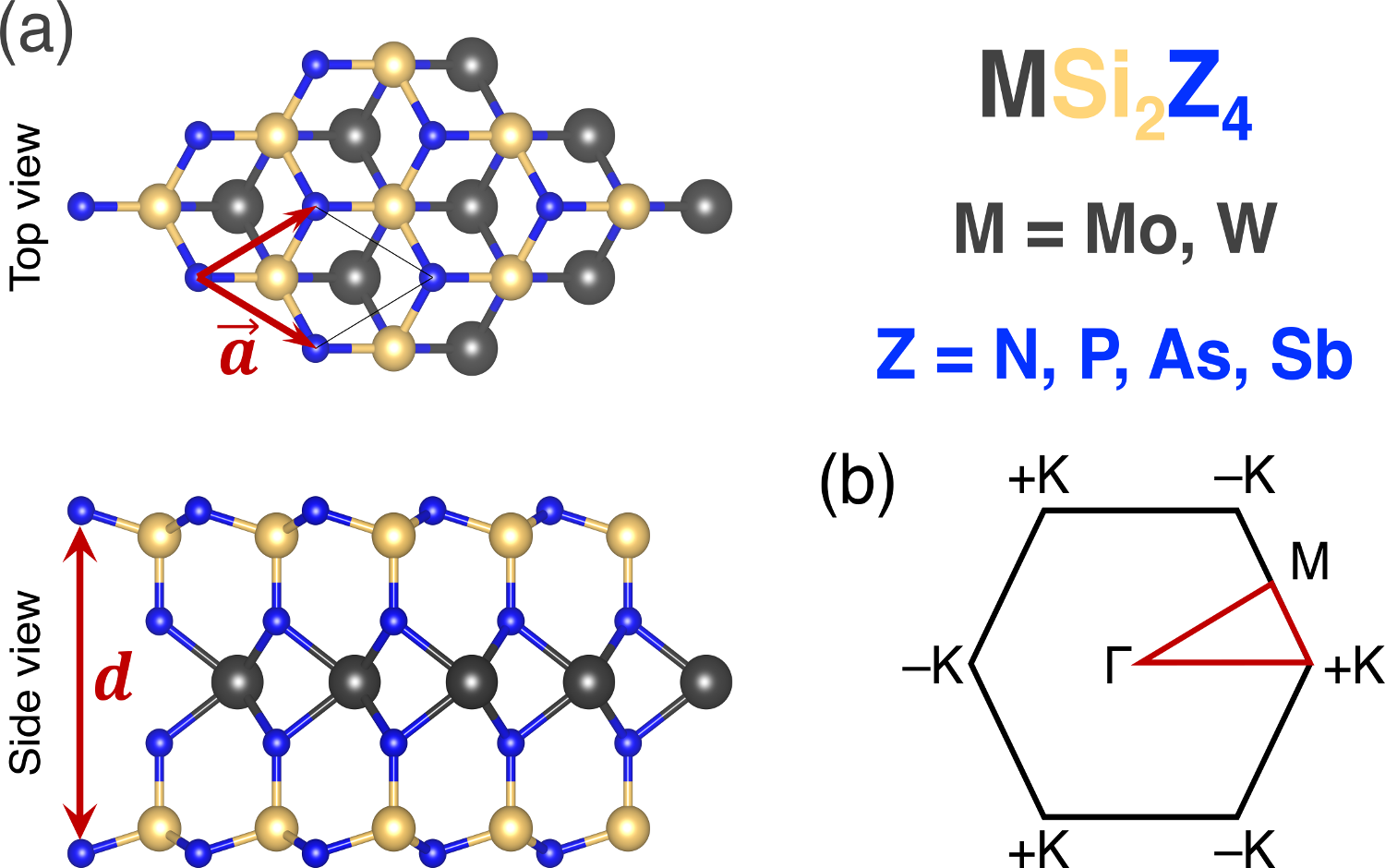}
 \caption{
 (a) Top and side views of 1L $M$Si$_2Z_4$ monolayer generic structure. Hexagonal unit cell is marked together with lattice vector $\vec{a}$ and the layer thickness $d$, defined as a distance between the outermost $Z$ atoms. 
 (b) The corresponding hexagonal BZ with the high-symmetry points and a path for band structure calculations (red lines). 
 }
\label{fig:structure} 
\end{figure*}

Subsequent density functional theory (DFT) calculations led to prediction of a large class of these new 2D vdW materials with general formula $MA_2Z_4$ (with $M$ - Group-2 or transition-metal elements, $A$ - Group-13 or -14 elements, and $Z$ - Group-15 or -16 elements.\cite{Wang2021,Novoselov2020a}
They share a common atomic structure consisting of 7 atomic layers, $Z-A-Z-M-Z-A-Z$ (see Figure~\ref{fig:structure}).
A 1L $MA_2Z_4$ can be viewed as a 1L $MZ_2$ sandwiched between two $AZ$ layers, or alternatively, a 1L $MZ_2$  intercalated into a 1L $A_2Z_2$.\cite{Wang2021}
Depending on the phases of the constituents (2$H$ or 1$T$ $MZ_2$ and $\alpha$ or $\beta$ $A_2Z_2$), the resulting hexagonal structure has $P\bar{6}m2$ space group ($D_{3h}$ point group), identical to 2$H$-MoS$_2$ analogues, or $P\bar{3}m1$ space group ($D_{3d}$ point group), identical to 1$T$-HfS$_2$ analogues.
Depending on the space group, as well as the number of valence electrons, 1L $MA_2Z_4$ can be non-magnetic or ferromagnetic semiconductors or metals, some of which exhibit topological and superconducting behavior.\cite{Wang2021}
Numerous properties of 1L $MA_2Z_4$ have been predicted by DFT calculations, including robust magnetic properties,
\cite{Cui2021,Chen2021,Li2021b}
defect-induced half-metalicity,\cite{Ray2021}
high piezoelectric coefficients,
\cite{Guo2020a,Guo2020,Guo2021a,Mortazavi2021,Guo2020b,Guo2021b} 
ferroelectricity switchable via interlayer sliding, \cite{Zhong2021a}
high intrinsic lattice thermal conductivity, \cite{Yu2021,Shen2022,Yin2021}
exceptional thermoelectric performance, \cite{Huang2020,Bafekry2021a,Guo2021}
anomalous spin and valley Hall effects, \cite{Wang2021,Li2021a,Cui2021,Zhou2021}
Ising superconductivity,\cite{Wang2021}
giant tunneling magnetoresistance,\cite{Wu2022}
tunable absorption coefficient,\cite{Jian2021} and effective chemical adsorption properties.\cite{Bafekry2021b,Cui2021a,He2022}  
Large on/off ratios were obtained in simulations of 1L $MA_2Z_4$ transistors. \cite{Sun2021,Zhao2021,Nandan2022,Ghobadi2022}
This makes 1L $MA_2Z_4$ promising candidates for device applications and stimulates further theoretical and experimental studies.

The class of 1L $M$Si$_2Z_4$ was theoretically predicted to have direct band gaps at the $K$ valleys; the only exceptions are MoSi$_2$N$_4$ and WSi$_2$N$_4$, for which the transition is between $\Gamma$ and $K$ points. Simulations suggest, however, that application of a moderate in-plane strain can shift it to the $K$ point.\cite{Li2020,Guo2021,Ai2021,Jian2021,BabaeeTouski2021,Wang2021c}
Similarly to 1L MoS$_2$ analogues, $K$ valleys of 1L $M$Si$_2Z_4$ host two excitons A and B, originating from transitions between spin-split $v\pm$ (valence) and $c\pm$ (conduction) bands. The spin ordering of bands and optical selection rules are identical to 1L MoS$_2$, where A and B transition at $\pm K$ valleys couple to $\sigma\pm$ polarized light and D transition couples to linearly polarized light.\cite{Zeng2012,Wang2018}
Additionally, in compounds with Z = P, As and Sb, another two conduction bands $c1\pm$ are present above $c\pm$. They are involved in two higher-energy excitonic transitions A* and B* with circular polarizations opposite to A and B at given $K$ valley, and a linearly polarized D* transition.

Despite numerous studies of electronic structure of 1L $M$Si$_2Z_4$, the literature on their quasi-particle band gaps and exciton binding energies is rather scarce. The recent GW+BSE calculations highlighted the band gap renormalization of more than 1.0~eV and exciton binding energy of several hundreds of meV.\cite{Sheoran2022,Tian2021,Yadav2022,Wu2021a,Zhong2022}
The convergence of GW band gap and exciton binding energy with respect to calculation parameters was carefully checked only in Refs.~\cite{Wu2021a,Zhong2022}, yielding indirect and direct gap of 2.82--2.86~eV and 3.09--3.13~eV, respectively, and exciton A binding energy of 0.63~eV in 1L MoSi$_2$N$_4$.
Furthermore, since $MA_2Z_4$ compounds are layered materials with band edges around the $K$ points, the well-established effective modelling of exciton physics in TMDC-based systems\cite{Berkelbach2013a,ChernikovPRL2014,Scharf2017PRL,Stier2018PRL,ChoPRB2018,Zollner2019,Zollner2019CrI3,Goryca2019,Zollner2020PRB,Henriques2020PRBtmdc,Lin2021a} can be applied here as well. This would allow additional understanding of substrate and encapsulation effects on the excitonic properties, which goes beyond the GW+BSE analysis of bare 1Ls.

Due to the similarities of lattice symmetry and electronic structures of 1L 2$H$-TMDCs and 1L $MA_2Z_4$, they are expected to exhibit similar optical response to external magnetic field, especially exciton $g$-factors or diamagnetic coefficients ($\alpha$).
The $g$-factors, which describe the linear (Zeeman) shifts of excitonic energies, help to assign the optical peaks to specific excitonic complexes, band structure transitions, and regions of heterostructures. \cite{Ciarrocchi2019,Seyler2019,Wozniak2020,Zinkiewicz2021} From the diamagnetic coefficients, which describe the quadratic energy shift, one can derive exciton reduced masses and radii.\cite{Rogers1986,Goryca2019,Chen2019} From the technological point of view, magnetoexcitonic properties are important for valleytronic devices. For instance, numerical studies of TMDCs quantum dots have shown that large $g$-factors are desirable to efficiently tune the spin-valley qubits at moderate external magnetic fields.\cite{Kormanyos2014,Goh2020,Pawlowski2021}
While exciton $g$-factors are widely investigated for TMDCs,\cite{Arora2021} such magnetooptical properties of 1L $MA_2Z_4$ have not been reported so far.

In this work, we present a comprehensive study of electronic and excitonic properties of 1L $M$Si$_2Z_4$ (shown in Figure~\ref{fig:structure}a) from DFT and model BSE (Bethe-Salpeter Equation) calculations. Chemical trends of band gaps and SOC splittings of bands at $K$ point were also analyzed. We report the binding energies and radii of excitons involving $v\pm$, $c\pm$ and $c1\pm$ states at $K$ valleys. We also provide the values of their Landé $g$-factors and diamagnetic coefficients, which describe the evolution of excitonic energies under magnetic fields. We predict large positive $g$-factors for excitons that involve higher conduction bands emerging in $Z$ = P, As and Sb compounds. Additionally, we observe a SOC-induced band inversion for the heaviest here studied material, WSi$_2$Sb$_4$, suggesting that this compound might host topological properties.

\section{Results and Discussion}
\subsection{\label{sec:geometry}Geometrical structure}

The optimized lattice constants ($a$) and slab thicknesses ($d$) are shown in Table~\ref{tab:geo_Eg_split} together with electronic properties, which will be discussed later. 
Both parameters and their ratio, $d/a$, monotonically increase with the mass of $Z$ element. Similarly to 2$H$-TMDCs, Mo and W compounds with the same $Z$ have almost identical lattice constants (their relative difference does not exceed 0.2\%), which enables to create nearly commensurate MoSi$_2Z_4$/WSi$_2Z_4$ heterobilayers with interlayer twist angles of 0$^\circ$ and 60$^\circ$. 

\setcounter{table}{0}
\makeatletter 
\renewcommand{\thetable}{\@arabic\c@table}
\makeatother
\begin{table*}[ht!]
 \caption{\label{tab:geo_Eg_split} Calculated lattice constants ($a$) and slab thicknesses ($d$); direct band gaps at $K$ ($E_g$) from PBE/HSE06 calculations, including spin-orbit coupling (indirect $\Gamma-K$ gaps of MoSi$_2$N$_4$ and WSi$_2$N$_4$ are given in parentheses); spin splittings of $v$, $c$ and $c1$ bands ($\Delta_{v}$, $\Delta_{c}$ and $\Delta_{c1}$) from PBE/HSE06 calculations. 
 }
 \begin{ruledtabular}
 \begin{tabular}{lccccccc}
 System & $a$ [\AA] & $d$ [\AA] & $E_g$ [eV] & $\Delta_{v}$ [meV] & $\Delta_{c}$ [meV] & $\Delta_{c1}$ [meV] \\ 
  \hline
 MoSi$_2$N$_4$  & 2.900 & ~7.003  & 2.021(1.787)/2.487(2.350) &  129/169 & 3/17 & - \\
 MoSi$_2$P$_4$  & 3.453 & ~9.347  & 0.621/0.864 & 137/202 & 4/30 & 87/112 \\
 MoSi$_2$As$_4$ & 3.600 & ~9.905  & 0.512/0.684 & 179/289 & 16/58 & 113/160 \\
 MoSi$_2$Sb$_4$ & 3.879 & 10.882  & 0.262/0.364 & 262/422 & 25/91 & 146/232 \\
 \hline
  WSi$_2$N$_4$  & 2.905 & ~7.014  & 2.162(2.107)/2.660(2.672) & 403/502 & 10/40 & - \\
  WSi$_2$P$_4$  & 3.457 & ~9.350  & 0.297/0.433 & 439/609 & 7/78 & 219/269 \\
  WSi$_2$As$_4$ & 3.603 & ~9.902  & 0.211/0.258 & 503/740 & 25/121 & 266/341 \\
  WSi$_2$Sb$_4$ & 3.885 & 10.879  & 0.032/0.186 & 560/953 & 32/186 & 306/423 \\
 \end{tabular}
 \end{ruledtabular}
\end{table*}

The stability of the considered compounds is confirmed by their phonon band structures, which show no imaginary phonon modes, see Figure~S1.
The phonon dispersion relation and element-projected density of states show distinct three regions: i) high-frequency dominated by N and Si atoms, ii) medium-frequency involving all elements, and iii) the low-frequency dominated by transition-metal and As or Sb. For the heavier layers, region i) has only contributions from silicon, whereas regions ii) and iii) merge together.
Overall, the three regions are separated by smaller or larger energy gaps.

\subsection{\label{sec:electronic}Electronic structure}

\begin{figure*}[ht!]
 \includegraphics[width=0.7\textwidth]{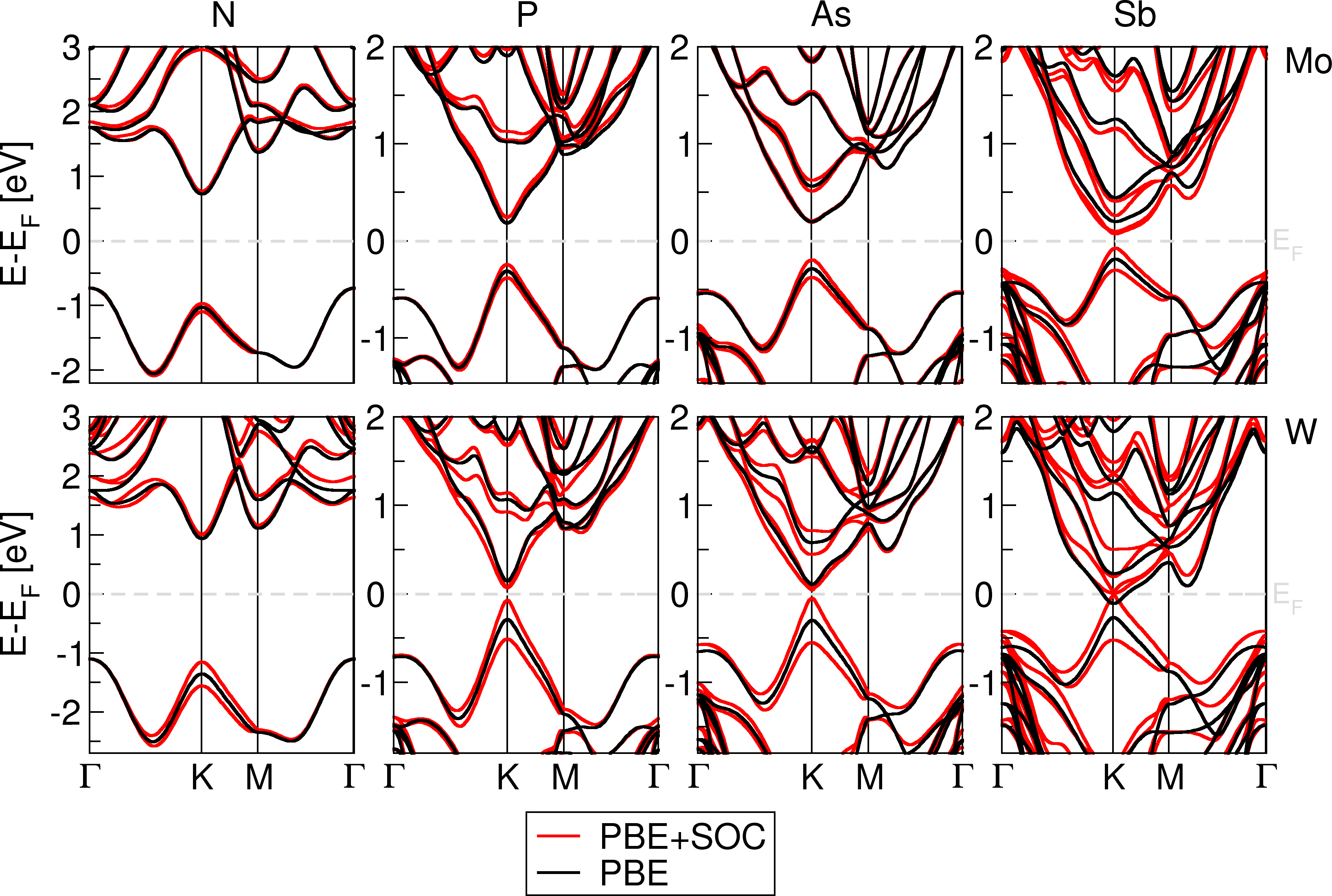}
 \caption{\label{fig:BS_PBE}Electronic band structures of all compounds calculated at the PBE level of theory without (black) and with SOC (red). (Top panel) and (bottom panel) show data for Mo- and W-based systems, respectively. The Fermi level was set to the middle of the band gap and shifted to zero for the PBE+SOC results. The black lines were shifted such that VBM at $K$ is in the middle of the (red) SOC spin-split VBM. Note different energy scales used for more clarity.}
 \end{figure*}

Electronic band structures of 1L $M$Si$_2Z_4$ calculated using PBE without and with SOC are shown in Figure~\ref{fig:BS_PBE}. The valence band maximum (VBM) of MoSi$_2$N$_4$ and WSi$_2$N$_4$ is located at $\Gamma$ point, 233~meV and 54~meV above the band extrema at $K$, respectively. For the other compounds, VBM shifts to the $K$ point, which is more than 200~meV higher than valence band at $\Gamma$. The conduction band minimum (CBM) is always located at the $K$ point, with secondary band extremum at or near $M$ point. The energetic separation between conduction band valleys at $K$ and $M$ points is the smallest for WSi$_2$N$_4$ (156~meV) and exceeds 200~meV for all other compounds. As a consequence, MoSi$_2$N$_4$ and WSi$_2$N$_4$ have indirect fundamental band gaps between $\Gamma$ and $K$ points, and all other compounds are direct gap semiconductors with transitions at $K$.
The direct gap materials have also another higher energy minimum in conduction band (CBM+1; $c1$) at $K$, which approaches the CBM with increasing the mass of $Z$. These additional minima result in higher-energy excitonic transitions (see Figure~\ref{fig:BS_PBE} and Section~\ref{sec:excitons}).

 The band structures and resulting bang gap values were also recalculated with a higher-level density functional, namely HSE06, and the results are shown in Figure~S2
 , along with PBE band structures.
 In all cases, HSE06 yields larger band gaps than PBE, as shown in Table~\ref{tab:geo_Eg_split} and Figure~S3.
 The main difference between both methods is that the VBM of WSi$_2$N$_4$ shifts to $K$, making it also a direct gap material.
 All the other chemical trends of band gaps are preserved (see Figure~S3).

The calculated band gap values are given in Table~\ref{tab:geo_Eg_split} and show the following general trends: the values of Mo-based compounds are generally larger than these of heavier W-based compounds with the same $Z$, except for nitrogen. When increasing the mass of $Z$, the band gaps decrease, with the largest reduction from N to P (see Figure~S3).
MoSi$_2$N$_4$ and WSi$_2$N$_4$ exhibit the largest gaps in the visible light range, while the other compounds have their gaps in the infrared regime.
 
SOC in non-centrosymmetric materials, as studied here, can lift the spin degeneracy of electronic bands, leading to spin-split bands along the $\Gamma$-$K$ and $K$-$M$ directions (see Figure~\ref{fig:BS_PBE}).
We define the spin splittings of $v$, $c$, and $c1$ bands at $\pm K$ points ($C_3$ point group) as $\Delta_v = |E(\Gamma_5) - E(\Gamma_4)|$,  $\Delta_c = |E(\Gamma_6) - E(\Gamma_5)|$, and  $\Delta_{c1} = |E(\Gamma_6) - E(\Gamma_4)|$, respectively, using the double group (with SOC) irreducible representations.
Their magnitudes are given in Table~\ref{tab:geo_Eg_split} and the resulting trends are shown in Figure~S4.
For both functionals, the spin ordering of bands is identical to the case of 1L MoS$_2$, with a crossing of $c+$ and $c-$ bands near the $K$ point.
The values are in good agreement with previous theoretical reports.\cite{Li2020}
Depending on the employed density functional, they are in the ranges of 130 -- 950~meV for $\Delta_v$, 3 -- 190~meV for $\Delta_c$, and 90 -- 420~meV for $\Delta_{c1}$.
As expected, the splittings are higher for W-based compound and monotonically increase with the mass of $Z$. 

At the HSE06 level, $\Delta_v$ in WSi$_2Z_4$ (with $Z$ = P, As, Sb) exceeds the largest measured value for 1L WSe$_2$ (513~meV \cite{Le2015}), one of the heaviest 2$H$-TMDC.
Furthermore, all compounds have significant $\Delta_c$ values, even larger than the only experimental value in the 1L WSe$_2$ (14~meV).\cite{Kapuscinski2021}
Such large spin splittings of bands would yield significant energetic resolution of excitonic peaks even at room temperature, as it was already shown for MoSi$_2$N$_4$.\cite{Hong2020a}.

The atomic-orbital decomposition of wave function on the bands structures of MoSi$_2$N$_4$ and WSi$_2$N$_4$ are shown in Figure~S5.
The band edges are dominated by states that originate from transition-metal atom $d$ orbitals. This orbital character is also shared by all the other systems (see Figures~S6 and S7).
A detailed analysis of orbital contributions is provided in Figure~S8.
The $v\pm$ states at the $K$ point are in $\approx$80\% composed of $d_2$ ($d_{xy}$ and $d_{x^2-y^2}$) orbitals of transition-metal atoms with small admixture of $p_0$ ($p_z$) and $p_1$ ($p_x$ and $p_y$) orbitals of pnictogenes. The $c\pm$ states are constructed in $\approx$90\% by $d_0$ ($d_{z^2})$ orbitals of transition-metal atoms with a small contribution of metal $s$ and $Z$ $p_1$ orbitals. The upper conduction states $c1\pm$ are still dominated by transition-metal atoms $d_2$ orbitals ($\approx$40-60\%), but with a significant contribution of other orbitals: $s$ and $p_0$ of Si, and $s$ and $p_1$ of pnictogenes.

It is worth noting that the heaviest of the considered compounds, WSi$_2$Sb$_4$, exhibits the largest SOC effects due to the presence of two heavy elements (cf. Figure~S9).
Examining the orbital composition of $v\pm$ and $c\pm$ bands, we observe a reordering of the spin-split bands $v+$ and $c+$ with respect to all the other compounds. Simultaneously, analysis of the wave function symmetry reveals a change of the irreducible representations of the considered bands. Such a feature is an indication of SOC-induced transition from a trivial to a topological state\cite{Kurpas2019,Cano2021} (more detailed analysis of the wave function symmetry and topological invariants of the considered bands is beyond the scope of present work).

\begin{figure*}[ht!]
 \includegraphics[width=0.6\columnwidth]{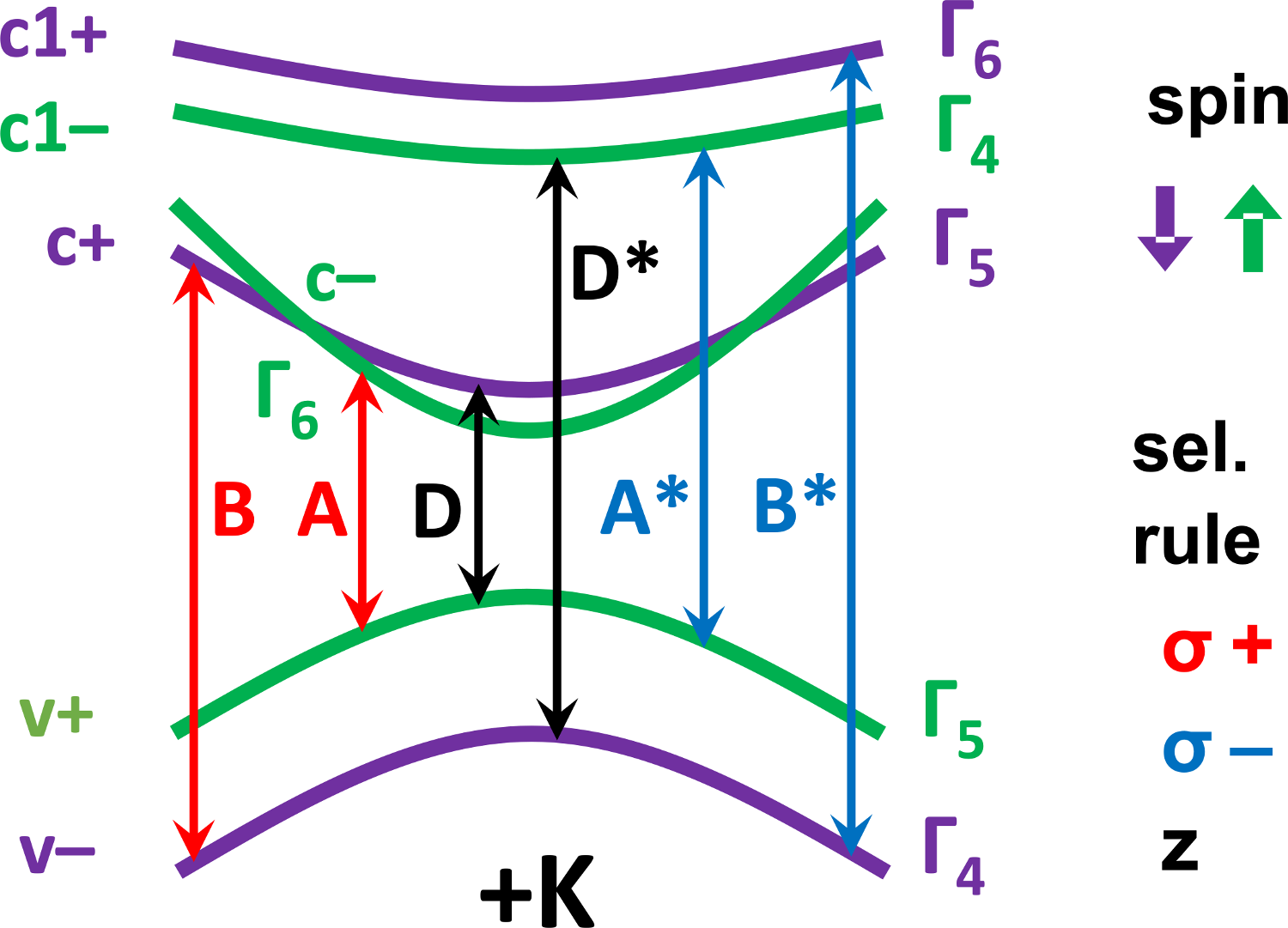}
 \caption{\label{fig:sel_rules} 
 Sketch of the electronic bands of 1L MSi$_2$Z$_4$ near the band gap at $+K$ valley in the presence of SOC. Higher ($v+$, $c+$ and $c1+$) and lower energy ($v-$, $c-$ and $c1-$) spin-splitted bands are coloured green for spin up and violet for spin down. The irreducible representations of band symmetries in double point group notation are provided. The optically active transitions are marked with arrows. Red, blue and black colours correspond to $\sigma+$, $\sigma-$ and $z$ polarization of light, respectively.
 }
\end{figure*}

The spin ordering of spin-split bands at $+K$ leads to specific valley optical selection rules, sketched in Figure~\ref{fig:sel_rules}: spin-conserving transitions A and B ($v\pm \leftrightarrow c\mp$) couple to $\sigma \pm$ light at $\pm K$ valleys. Interestingly, transitions to the upper conduction band $c1$ ($v\pm \leftrightarrow c1\mp$), which we refer to as A* and B*, have opposite selection rules -- they couple to $\sigma\mp$ light at $\pm K$.
This means that a photon with adequate energy and circular polarization can selectively populate $c\pm$ or $c1\pm$ conduction states in $\pm K$ valleys. This multiple-folded valley property makes 1L $M$Si$_2Z_4$ promising candidates for multiple-information processing.\cite{Yang2020}.
Additionally, there are two $z$-polarized transitions D and D* ($v+ \leftrightarrow c+$ and $v- \leftrightarrow c1-$). The remaining two transitions, $v- \leftrightarrow c1-$ and $v+ \leftrightarrow c1+$, are optically inactive.
WSi$_2$Sb$_4$ exhibits distinct selection rules, as a consequence of aforementioned band inversion (see Figure~S9).

The complete list of all the transition energies from PBE and HSE06 simulations, together with their optical polarizations and dipole strengths from PBE calculations, are provided in Tables~S1--S8.
Transitions A and B have the largest dipole strengths, compared with A* and B* which are one order of magnitude less intense, but should also be observable in optical reflection measurements and modulation spectroscopy techniques, like electro- and photoreflectance.
On the other hand, the A* and B* energies are approximately twice larger than for A and B.
This, along with their selection rules opposite to A and B, makes the $Z$ = P, As, and Sb compounds promising candidates to study quantum interference effects, like electromagnetically induced transparency, recently reported for 1L and twisted bilayer WSe$_2$.\cite{lin2019quantum,Lin2021,Lin2021a} Therefore, a deeper insight into the properties of A* and B* transitions in such materials is desirable.
The $z$-polarized transitions D and D* are 3--4 orders of magnitude less intense, similar to 1L 2$H$-TMDCs.\cite{FariaJunior2022}
The dipole strengths of A, B and D (A*, B* and D*) transitions decrease (increase) with the mass of $Z$ and are generally larger for W-based compounds.

It is worth noting that the HSE06 calculated band gaps of MoSi$_2$N$_4$ (direct 2.35~eV and indirect 2.49~eV), as well as the energy of B transition and VBM spin splitting at $K$ of 2.67~eV and 190~meV, respectively, are in decent agreement with values of 1.94~eV, 2.21~eV, 2.35~eV, and 140~meV measured at room temperature for this compound.\cite{Hong2020a}
On the other hand, the only converged GW+BSE calculations available in the literature yield 2.5~eV energy of exciton A,\cite{Wu2021a} almost identical to our HSE06 result, which does not take into account many-body effects and exciton binding energy. It indicates that HSE06 may give a good estimate and chemical trends of band gaps and spin splittings also for other compounds of 1L $MA_2Z_4$ family, at a moderate computational cost. 

\begin{figure*}[h!]
 \includegraphics[width=0.7\textwidth]{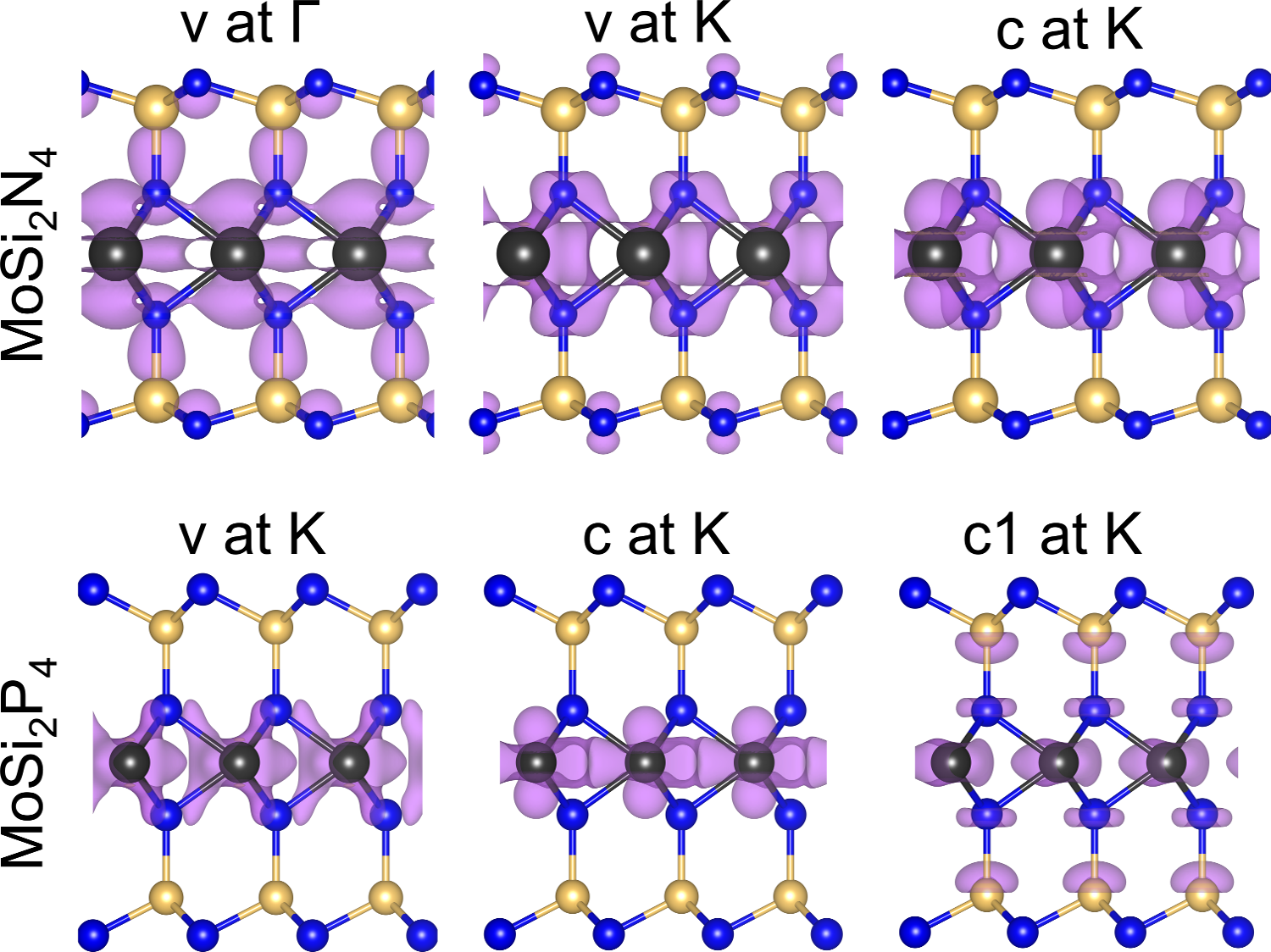}
 \caption{\label{fig:PCD_MSi2N4} Electron and hole states of 1L MoSi$_2Z_4$ (Z = N and P) as exemplary materials at the relevant valleys and their wave function extent in real space - isosurfaces of the partial charge density with isovalue of 0.007~e~\AA$^{-3}$. v - valence band maximum, c - conduction band minimum, c1 - higher energy conduction band at $K$.}
\end{figure*}

Next, we calculated and plotted the real-space wave function at the relevant $k$-points. Figure~\ref{fig:PCD_MSi2N4} shows exemplary wave function plots for 1L MoSi$_2$N$_4$ and MoSi$_2$P$_4$ (see Figures~S10 and S11
for all other compounds).
In general, the electron and hole states are situated at the $K$ point, except for MoSi$_2$N$_4$, in which the hole states are at $\Gamma$.
The wave function plots show the localization of states at $K$ to the middle part of the layer, mostly on the transition-metal atom and, for lighter materials, on the inner $Z$ atoms.
Increasing the mass of $Z$, the localization increases and eventually it is dominated by just the metal atom for Sb-based compounds.
Since the $M$ atom layer is situated in the middle of a 7-atomic-layer slab, the electron and hole states are effectively screened inside the monolayer.
Yao et al.~\cite{Yao2021} showed also that this behaviour persists for bilayers, showing layer-independent electronic properties (e.g., band gaps).
This is different for 1L MoSi$_2$N$_4$, where hole states are delocalized to all atomic layers. This is also true for the higher-energy electron states ($c1\pm$).
Such differences in the carrier state localization may result in different exciton lifetimes.

While the wave function plots might not be of great interest on their own, they give an indication of stronger spatial separation in case of type-II band alignment in heterostructures (work in progress, but beyond the scope of this paper).\cite{Pham2021,Zhao2022}

\subsection{\label{sec:excitons}Excitonic properties}

In order to host a stable excitonic state, the dispersion of its electron and hole band energy difference $E_m(\mathbf{k})-E_n(\mathbf{k})$, ($n=v\pm$, $m=c\pm$, $c1\pm$) must have a local minimum in the reciprocal space. We used the calculated band structures to check that this condition is fulfilled for all the considered transitions in all compounds, see Figures~S12 and S13.
The band effective masses and screening lengths $r_0$ of 1L MoSi$_2Z_4$, provided in Table~S1, as obtained from PBE calculations, are used in model BSE calculations, as described in Section~\ref{sec:methods}.
The reduced masses $m_r$ (see Tables~S2--9) of most of the excitons are similar to the case of 1L TMDCs,\cite{Arora2021} except for WSi$_2Z_4$, $Z$ = P, As, Sb, where they may be lower than 0.1.
The values of $r_0$ are increasing with the decrease of band gap and are similar to or exceed the typical values of 1L 2$H$-TMDCs.\cite{Berkelbach2013a} The resulting exciton binding energies, $E_B$, are provided in Tables~S2--S9.
The $E_B$ of all excitons in MoSi$_2$N$_4$ and WSi$_2$N$_4$ are of the order of 450--500 meV, with the highest values for indirect excitons, as a consequence of their low reduced masses. They are comparable to the binding energies observed in 1L 2$H$-TMDCs,\cite{Wang2018} which should ensure the stability of excitons against thermal dissociation. The values for A*, B* and D* excitons are similar or slightly larger than for A, B, and D, and show a systematic decrease, along with the decreasing band gap for heavier $Z$ atoms. 

\begin{figure*}[ht!]
 \includegraphics[width=0.7\textwidth]{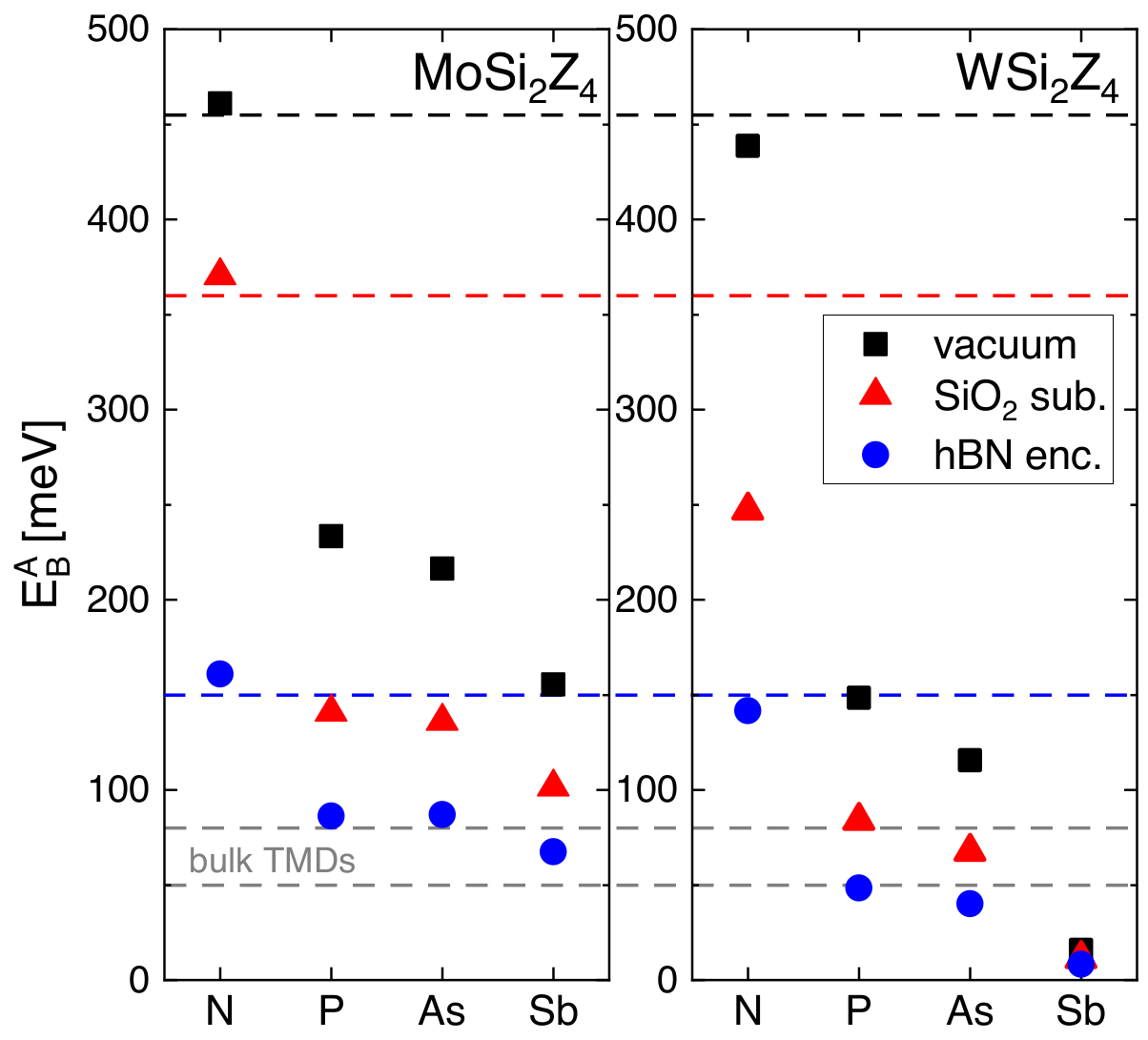}
 \caption{\label{fig:EbA} Exciton binding energy $E_B$ of A exciton in 1L $M$Si$_2Z_4$ compounds for different dielectric environments: in vacuum ($\varepsilon = 1$), on SiO$_{\textrm{2}}$ substrate ($\varepsilon = 2.4$) and encapsulation in hBN ($\varepsilon = 4.5$). For comparison, the corresponding $E_B$ values of A exciton are marked with color dashed lines for 1L WSe$_2$ and with grey dashed lines for Group-6 bulk TMDCs.\cite{Birowska2021}}
\end{figure*}

The binding energy of an exciton is strongly sensitive to its dielectric environment. Model BSE enables to study this dependence by introducing an effective relative dielectric constant ($\epsilon$) of the systems' surroundings (see Section~\ref{sec:methods}). As shown in Figure~S14
, $E_B$ decreases with increasing $\epsilon$, as in case of 1L 2$H$-TMDCs. Figure~\ref{fig:EbA} presents this dependence for A exciton in all the considered compounds in three typical situations: in vacuum ($\varepsilon = 1$), on SiO$_{\textrm{2}}$ substrate ($\varepsilon_{b} = 3.9$ \cite{SiO2}, $\varepsilon = 2.4$; see Section~\ref{sec:methods} for details)  and encapsulated in hBN ($\varepsilon = 4.5$ \cite{hBN}), compared to the corresponding values of 1L WSe$_2$ and Group-6 bulk TMDCs.\cite{Birowska2021} As expected, $E_B$ is significantly smaller for SiO$_2$ substrate or hBN-encapsulation, compared with the vacuum value.

For N-based compounds, the binding energies are comparable to the case of 1L WSe$_2$, which exhibits the largest $E_B$ of 1L 2$H$-TMDCs. The values gradually decrease with heavier $Z$ atoms, for WSi$_2$P$_4$ and WSi$_2$As$_4$ reaching the range of Group-6 bulk TMDCs. The binding energy of WSi$_2$Sb$_4$ is below the thermal energy at room temperature. It should be noted that our $E_B$ of A exciton from model BSE for MoSi$_2$N$_4$ (0.46~eV) is in a decent agreement with GW+BSE calculations of Wang et al.~\cite{Wu2021a} (0.63~eV). This discrepancy can be attributed to the different values of effective masses and the screening length obtained within different approaches.

\subsection{\label{sec:gfactors}Exciton $g$-factors and diamagnetic coefficients}

\begin{figure*}[h!]
 \includegraphics[width=0.7\columnwidth]{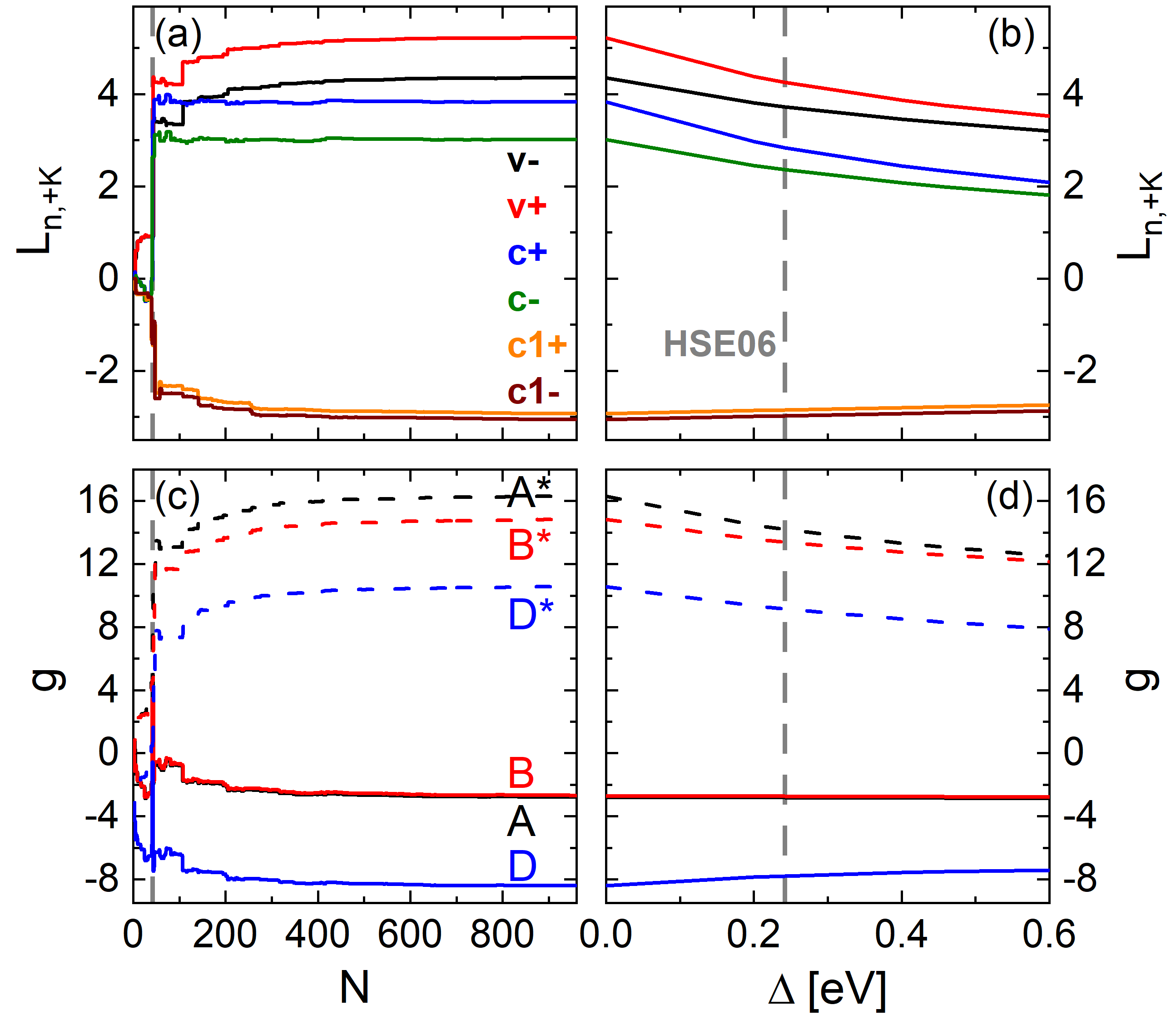}
 \caption{\label{fig:g-convergence}  Impact of basis set size $N$ and scissor correction $\Delta$ on orbital angular momenta and exciton $g$-factors in 1L MoSi$_2$P$_{\textrm{4}}$. (a) Convergence of orbital angular momenta $L_{n,\textrm{+K}}$ of the two highest valence band states $(n = v\pm)$ and four lowest conduction band states $(n = c\pm$, $c1\pm)$ at the $+K$ point with respect to the number of bands $N$ included in the summation of Equation~(\ref{eqn:Lnk}). (b) Convergence of the $g$-factors of A, B, D, A*, B* and D* excitons. $N$ = 1 is the lowest-energy state of the valence shell and the valence band maximum is indicated by a dashed vertical line. (c) Impact of the scissor correction $\Delta$ on the orbital angular momenta $L_{n,+K}$ and (d) the exciton $g$-factors. $\Delta=0$ corresponds to the band gap obtained from PBE simulations. The dashed vertical line indicates the band gap calculated at the HSE06 level of theory. }
\end{figure*}

After analyzing the electronic structure and exciton binding energies, we focused on the magnetooptical properties of 1L $M$Si$_2Z_4$.
First, the convergence of orbital angular momenta of bands $L_{n,+K}$ and exciton $g$-factors with respect to the number of states involved in the summation in Equation~(\ref{eqn:Lnk}) must be performed (see Figures \ref{fig:g-convergence}(a) and (b) for exemplary MoSi$_2$As$_4$ results).
The largest contribution to $L_{n,+K}$ is at the band gap (dashed vertical line). Around 800 bands are required to converge the $g$-factors to the precision of 0.1.
This is more than in the case of 1L TMDCs, where around 300--500 bands are required.\cite{Wozniak2020,Forste2020,FariaJunior2022}
The converged values of orbital angular momenta for all compounds are given in Table~S10.
We observe that $L_{n,+K}$ for $v\pm$ and $c\pm$ are positive and of the same order as for TMDCs. Interestingly, $L_{n,+K}$ of $v+$ and $c+$ in WSi$_2$P$_4$ and WSi$_2$As$_4$ are much larger than in the other compounds. This is due to a large contribution of $p_{cv}$ in the summation of Equation~(\ref{eqn:Lnk}).
The $L_{c1\pm,+K}$ values for $Z$ = P, As, Sb are negative. For VBM at $\Gamma$, the orbital angular momentum is smaller than 0.01 in MoSi$_2$N$_4$ and smaller than 0.03 in WSi$_2$N$_4$.
The $g$-factors of A and B excitons are negative and significantly smaller in magnitude than these for 1L TMDCs, where values of around -4 are observed.\cite{Wozniak2020,Deilmann2020,FariaJunior2022} This indicates that simplistic models, which are based on orbital compositions of band edges and their effective masses, cannot accurately predict $g^{A,B}$ in $M$Si$_2Z_4$. The $g$-factor of D exciton is lower from $g^{A,B}$ by more than -4, which is due to the spin-flip contribution. The $g$-factors of A*, B* and D* excitons are positive and have much larger magnitude than for A, B and D. The former origins from the inverted selection rules ($\sigma\pm$ light couples to $\mp K$ valleys). The latter is caused by positive $L_{c1\pm,+K}$. As a consequence, the Zeeman splittings for A* and B* excitons are approximately 4--5 times larger than for A and B in $M$Si$_2Z_4$  and 1L TMDCs. To the best of our knowledge, these are the largest values reported for hexagonal monolayers so far. This allows to achieve a sizeable spin and valley splitting of A* and B* based states in MoSi$_2Z_4$ quantum dots at lower magnetic fields than in the case of TMDCs.\cite{Kormanyos2014} 

As the standard DFT calculations with PBE functional severely underestimate the electronic band gaps, the resulting $L_{n,\mathbf{k}}$ are usually overestimated, due to the energy denominator in Equation~(\ref{eqn:Lnk}). This can be corrected on the easiest possible level by applying the scissor shift $\Delta$ to the conduction bands: $\varepsilon_{n,\mathbf{k}}' = \varepsilon_{n,\mathbf{k}} + \Delta$ for $n \geq c-$.
Figures \ref{fig:g-convergence}(c) and (d) present the influence of $\Delta$ on the orbital angular momenta and exciton $g$-factors. When increasing the band gap of MoSi$_2$As$_4$ to HSE06 value, $L_{n,\mathrm{k}}$ for $v\pm$, $c\pm$ significantly decrease with scissor, while the values for $c1+$ and $c1-$ slightly increase. As a consequence, the magnitudes of excitonic $g$-factors decrease. The largest slope is observed for A*, B* and D* excitons, while the values for A and B are nearly constant. The trends for A, B and D excitons are similar to the case of 1L TMDCs.\cite{Wozniak2020,FariaJunior2022} 

\begin{figure*}[h!]
 \includegraphics[width=0.7\columnwidth]{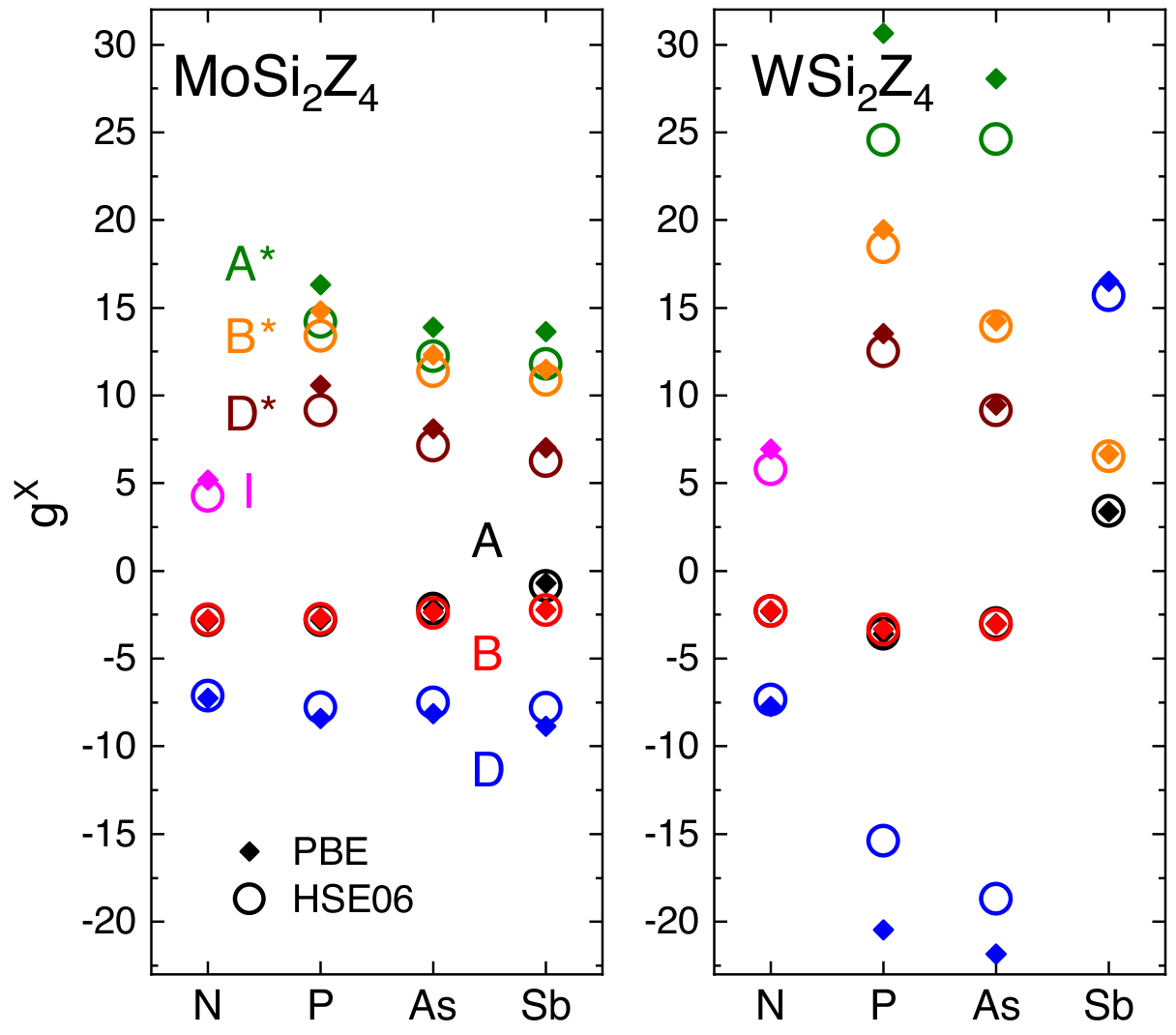}
 \caption{\label{fig:g-trends} Exciton $g$-factors, $g^X$, of $X$ = I, A, B, D, A*, B* and D* excitons in 1L $M$Si$2Z_4$ compounds calculated using band gaps from PBE (diamonds) and HSE06 (circles) simulations.}
\end{figure*}

The $g$-factors of all the considered excitons are shown in Figure~\ref{fig:g-trends}. Overall, $|g^X|$ are larger for W- than for Mo-based materials. As already discussed, for a given compound, the $g$-factors of A and B excitons are similar and do not depend on scissor correction. Increasing the mass of $Z$ element, their magnitudes decrease (increase) for materials containing Mo (W). The splitting $g^A-g^B$ increases with the mass of $Z$ and becomes sizeable in MoSi$_2$Sb$_4$.
For all compounds, except WSi$_2Z_4$, $Z$ = P, As, Sb, the $g$-factors of exciton D are in the range of -7 -- -9. The $g$-factors of A*, B*, and D* are always positive and reduce with increase in mass of $Z$ element. Their splitting $g^{A*}-g^{B*}$ is significantly larger than for A and B.
As in the case of MoSi$_2$As$_4$, the influence of scissor correction is more prominent for I, D, A*, B*, and D* than for A and B excitons. The effect is noticeably large for D and A* in WSi$_2$P$_4$  and WSi$_2$As$_4$, which stems from a strong reduction of $L_{n,+K}$ for $v+$ and $c+$ by scissor correction (see Tables~S7 and S8).

In addition to zero-momentum excitons, based on direct transitions at $K$ valleys, in MoSi$_2$N$_4$ and WSi$_2$N$_4$, we consider also finite-momentum excitons $I$, connected to indirect fundamental band gaps (VBM at $\Gamma$ and CBM at $K$) of these compounds. Their values are dominated by $L_{c+,+K}$, as $L_{v\pm,\Gamma} \approx 0$. Consequently, $g^I < g^{A,B}$ and $|g^I|$ decreases with the scissor correction. 

The $g$-factors in WSi$_2$Sb$_4$ do not follow the common trend. In contrast to the other materials, $g^A$ and $g^D$ ($g^B$ and $g^{A*}$) are positive (negative), which stems from the modified selection rules and strongly different $L_{n,+K}$ for $v-$, $c\pm$ and $c1+$, which are additionally almost insensitive to the scissor correction (see Table~S10).
This is related to the SOC-induced band inversion in $K$ valleys and requires further investigations, which are beyond the scope of this work. 

Finally, the quadratic part of exciton energy shift in magnetic field is quantified by the diamagnetic coefficient. In case of 1L TMDCs, where values $\alpha$ are lower than 1~$\mu$eV T$^{-2}$, very high magnetic fields are required ($B>$20~T) to observe the diamagnetic shifts.\cite{Goryca2019} We predict the values of diamagnetic coefficients in MoSi$_2Z_4$ basing on exciton reduced masses $m_r$ from DFT and expectation values of exciton radii $\braket{r^2}$ from BSE calculations, respectively. The values of $\alpha$ and $\sqrt{\braket{r^2}}$ are provided in Tables~S2--S9.
The exciton radii generally increase with the mass of $Z$ element. Consequently, diamagnetic coefficients also increase, reaching few or few tens of $\mu$eV T$^{-2}$ for excitons with low reduced mass and large $\sqrt{\braket{r^2}}$ in WSi$_2Z_4$, $Z$ = P, As, and Sb. Such large values of $\alpha$ might allow to observe significant diamagnetic shifts even at low magnetic fields.

\section{Conclusion}

We have investigated electronic properties of selected 1L $M$Si$_2Z_4$ employing first-principles calculations. Considered materials are direct gap semiconductors with transition at the $K$ point (except for $M$Si$_2$N$_4$, for which the fundamental transition is between $\Gamma$ and $K$). They also show emerging upper conduction bands $c1$, resulting in higher-energy excitonic transitions.
The electron and hole wave functions are mostly confined to the transition-metal atoms plane and the localization increases with the mass of $Z$ elements. The band gaps at $K$ point span a range of 2.66--0.19~eV and the spin splittings of valence, conduction, and upper conduction bands range from about 20~meV up to nearly 1~eV.
We have shown that the upper conduction states give rise to higher energy direct transitions with finite dipole strengths and circular optical selection rules.

We have studied the excitonic properties of 1L $M$Si$_2Z_4$ from the effective Bethe-Salpeter equation. Exciton binding energies and radii in range of 0.016--0.50~eV and 12--120~\AA, respectively, are predicted. We provide the dependence of binding energies on the dielectric constant, highlighting the typical experimental conditions.
We have investigated the magnetooptical properties of considered excitons, quantified by their Lande $g$-factors and diamagnetic coefficients.
The $g$-factors of band edge excitons are similar to the case of 1L 2$H$-TMDCs, while the values for higher energy transitions are positive and range from 6 to 30. Such large values have never been observed in hexagonal 1Ls before. The calculated diamagnetic coefficients also reach high values, up to 84~meV T$^{-2}$. Combining both, significant excitonic energy shifts and splittings are expected under relatively low magnetic fields.
Finally, we observed a SOC-induced band inversion in WSi$_2$Sb$_4$, which might indicate its topological nature and leads to peculiar values of its magnetooptical coefficients.

Thanks to their electronic and excitonic properties, 1L $M$Si$_2Z4$ compounds are important new members in 2D materials family, as they offer a number of robust excitons with significant magnetooptical responses. This makes the studied compounds and their layered heterostructures an attractive platform for studying fundamental spin-valley physics and potential use in optoelectronic, spintronic, and valleytronic devices. In these terms, 1L $M$Si$_2Z_4$ can be considered as complementary to 1L 2$H$-TMDCs, and desire further experimental and theoretical studies.

\section{Experimental Section}
\label{sec:methods}
\textbf{Computational Details}

All DFT calculations were carried out using the Vienna ab-initio simulation package (VASP).\cite{Kresse1996}
 The projector augmented wave (PAW) technique was used to describe the ionic potentials. For the geometry optimization, we employed the generalized gradient approximation (GGA) of the exchange correlation-functional within Perdew-Burke-Ernzerhof (PBE) parametrization\cite{Perdew1996} together with the D3 dispersion correction for vdW interaction of Grimme.\cite{Grimme2010} The plane-wave cutoff energy and the total energy convergence criterion were set to 500~eV and 10$^{-8}$~eV, respectively. The geometrical structures were fully optimized until the interatomic forces and stress tensor components were lower than $10^{-4}$~eV~\AA$^{-1}$ and 0.1~kbar, respectively. A vacuum region of at least 15~\AA\ was used to avoid spurious interactions between the periodically repeated layers. The BZ integrations were performed on a $\Gamma$-centered 12$\times$12$\times$1 Monkhorst-Pack $k$-grid with a Gaussian smearing of 0.1~eV. PBE and Heyd-Scuseria-Ernzerhof (HSE06) hybrid functional\cite{HSE06} were adopted to calculate the electronic band structures. SOC was taken into account during geometrical optimization and electronic structure calculations.
 The irreducible representations of electronic bands at $+K$ points were determined from the analysis of wave function symmetries. The density functional perturbation theory approach was used to calculate phonon dispersion spectra as implemented in PHONOPY code\cite{Phonopy} with a 4$\times$4$\times$1 supercell.
 The partial charge densities (PCD) were calculated using PBE functional. Structural illustrations and the band decomposed PCD plots in real space were created using the 3D visualization package VESTA.\cite{VESTA}
 
To compute the exciton binding energies and wave functions, we used the effective Bethe-Salpeter equation (BSE)\cite{Rohlfing2000PRB, FariaJunior2019PRB, Zollner2019} with the electron-hole interaction mediated by Rytova-Keldysh (RK) potential.\cite{Rytova1967MUPB, Keldysh1979JETP} The input parameters required for the effective BSE are: (i) the effective masses of the relevant conduction and energy bands and (ii) the parameters for the RK potential, namely, the screening length of the 2D material, $r_0$, and the effective relative dielectric constant of the surrounding media, $\varepsilon = (\varepsilon_t + \varepsilon_b)/2$, with $\varepsilon_{t(b)}$ being the dielectric constant of the top (bottom) material. Particularly, the effective masses of electrons and holes were obtained from quadratic fits to the band edges near $K$ and $\Gamma$ points, while the screening length was calculated using the in-plane components of static dielectric tensors including the ionic contributions (following Ref.~\cite{Berkelbach2013a}). These parameters are given in Table~S1.
of the Supplementary Material (SM) for all studied compounds. Finally, the effective BSE is solved numerically on a 2D $k$-grid sampled with 121$\times$121 $k$-points in a square region with sides ranging from $-0.5$ to 0.5~$\textrm{\AA}^{-1}$, yielding a $k$-point spacing of $\Delta k = 1 / 120 \; \textrm{\AA}^{-1}$ along each direction. Furthermore, to improve the convergence, the RK potential was averaged in a 2D $k$-grid submesh of 121$\times$121 points covering an area of $\Delta k^2$. 

The dependence of exciton energy on the applied external out-of-plane magnetic field is described as $E^X(B) = g \mu_B B + \alpha B^2$, where $g$ is the exciton $g$-factor, $\mu_B$ is the Bohr magneton, $B$ is the magnetic field induction, and $\alpha$ is the diamagnetic coefficient. The linear term is referred to as Zeeman shift and the quadratic term as diamagnetic shift. Experimentally, $g$ is conveniently determined from a linear fit to $E_{\sigma+}(B) - E_{\sigma-}(B)$ and $\alpha$ from a quadratic fit to $1/2(E_{\sigma+}(B) + E_{\sigma-}(B))$.

The exciton $g$-factors were calculated within the bands summation method described in Ref.~\cite{Wozniak2020}. The linear shift of band energy $n$ at point $\mathbf{k}$ under external out-of-plane magnetic field $B$ is quantified by a $g$-factor of a Bloch state $|n \mathbf{k} \rangle$ expressed as:
\begin{align}
g_{n,\mathbf{k}} = L_{n,\mathbf{k}} + S_{n,\mathbf{k}},
\label{eqn:gnk}
\end{align}
where $L_{n,\mathbf{k}}$ and $S_{n,\mathbf{k}}$ are the $z$-components of orbital and spin angular momenta, respectively. The orbital angular momentum is evaluated from a summation formula:\cite{Roth1959,Chang1996}
\begin{equation}
L_{n,\mathbf{k}} = \frac{1}{i m_0} \sum_{m=1, m \neq n}^N \frac{p^x_{nm,\mathbf{k}} p^y_{mn,\mathbf{k}} - p^y_{nm,\mathbf{k}} p^x_{mn,\mathbf{k}}}{\varepsilon_{n,\mathbf{k}} - \varepsilon_{m,\mathbf{k}}},
\label{eqn:Lnk}
\end{equation}
where $m_0$ is the rest mass of the electron, $\varepsilon_{n,\mathbf{k}}$ is the Bloch state energy, $p^{x(y)}_{nm,\mathbf{k}}$ are the momentum operator matrix elements between $n$ and $m$ states in $x(y)$ directions, and the summation runs over all $N$ states in the basis set.
For the relevant low-energy bands, we considered $S_{n,+K} = \pm1$.
The Zeeman splitting of exciton $X$ = A, B, D, A*, B* and D* is then described by excitonic $g$-factor:
\begin{align}
g^X = \pm 2 (g_{m,+K} - g_{n,+K}),
\label{eqn:g}
\end{align}
where $n=v\pm$, $m=c\pm$,  $c1\pm$, and the sign depends on optical selection rules at $+K$ point. Determination of sign of the indirect exciton $g$-factor requires analysis of the phonon-mediated optical selection rules\cite{Blundo2022PRL}, which is beyond the scope of our work. Therefore, we provide the absolute value $|g^I| = 2 |(g_{c-,\pm K} - g_{v+,\Gamma})|$.

Diamagnetic coefficients of direct excitons were evaluated as $\alpha = \frac{e^2}{8m_r} \braket{r^2}$, where $e$ is the elementary charge, $m_r = \frac{m_m m_n}{m_m + m_n}$ is the exciton reduced mass calculated from electron and hole effective masses $m_m$ and $m_n$, respectively, and $\braket{r^2}$ is the expectation value of the squared exciton radius in real space.
The exciton radii were extracted from the exciton wave functions calculated with the effective BSE.

\medskip
\textbf{Acknowledgements} \par 

We thank Kai-Qiang Lin, Maciej Molas, Maciej Bieniek, Jarosław Pawłowski, Thomas Brumme and Florian Arnold for stimulating discussions. TW, UA, MSR, AK thank the Deutsche Forschungsgemeinschaft (project GRK 2247/1 (QM3) and project CRC1415, number 417590517) for financial support and the high-performance computing center of ZIH Dresden for computational resources. TW acknowledges the financial support of National Science Centre, Poland within Project No. 2021/41/N/ST3/04516. AK acknowledges association with priority program (project SPP2244 (2DMP)). PEFJ acknowledges the financial support of the Deutsche
Forschungsgemeinschaft (DFG, German Research Foundation) SFB 1277 (Project-ID 314695032, projects B07 and B11) and SPP 2244 (Project No. 443416183). M.S.R acknowledges financial support by project SMART, financed by the Volkswagen Foundation as part of the program “Niedersachsisches Vorab - Digitalisierung in den Natururwissenschaften. 

\medskip

\setcounter{figure}{0}
\setcounter{table}{0}

\makeatletter 
\renewcommand{\thefigure}{S\@arabic\c@figure}
\renewcommand{\thetable}{S\@arabic\c@table}
\makeatother


\clearpage
\section{Supplementary Material}
\subsection{Supplementary Figures}


 \begin{figure}[ht!]\centering
\centering
 \includegraphics[width=1\textwidth]{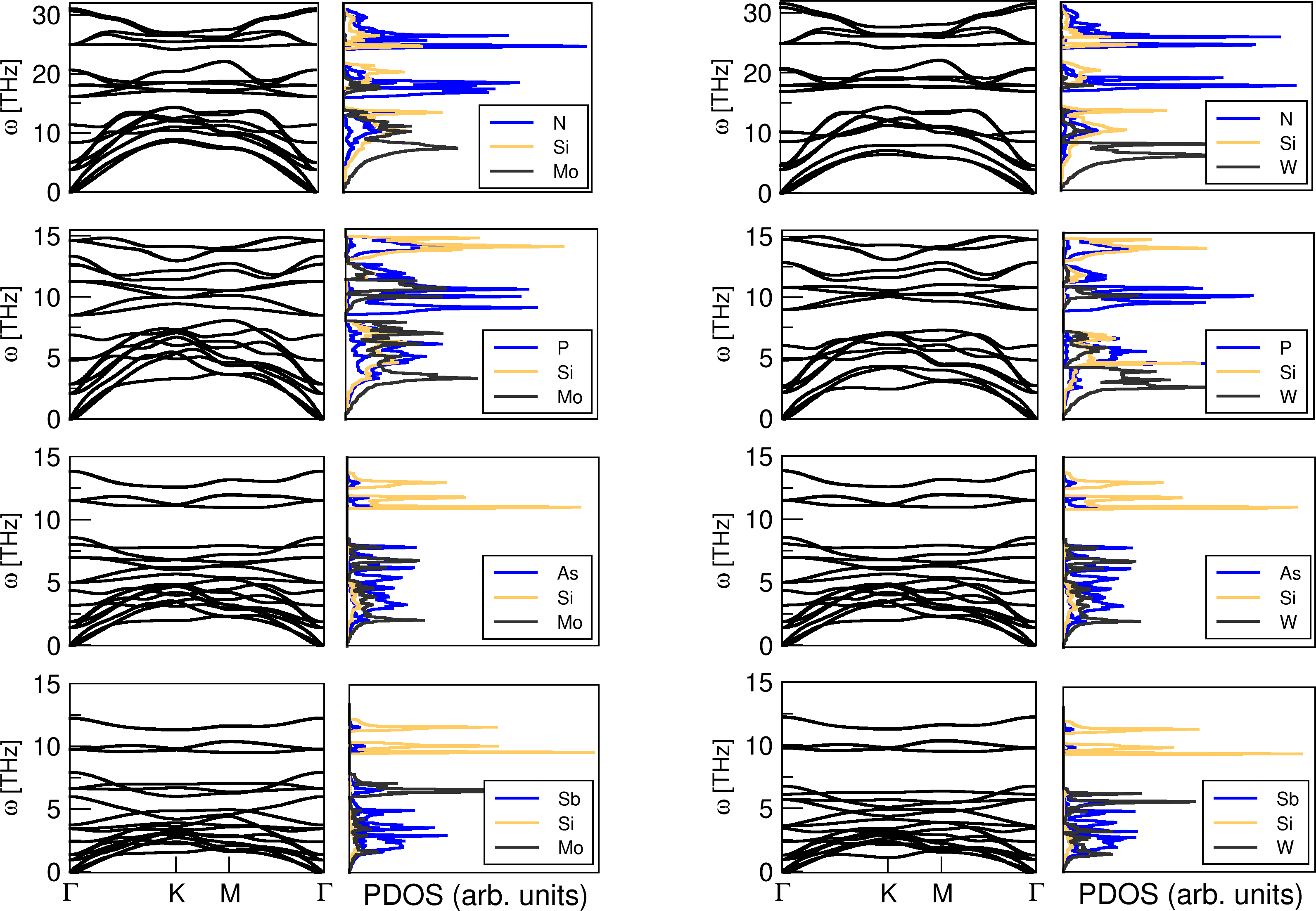}
 \caption{\label{fig:phonons} Phonon dispersion relations and densities of states (PDOS) of all studied compounds.}
\end{figure}


  \begin{figure}[ht!]\centering
\includegraphics[width=\textwidth]{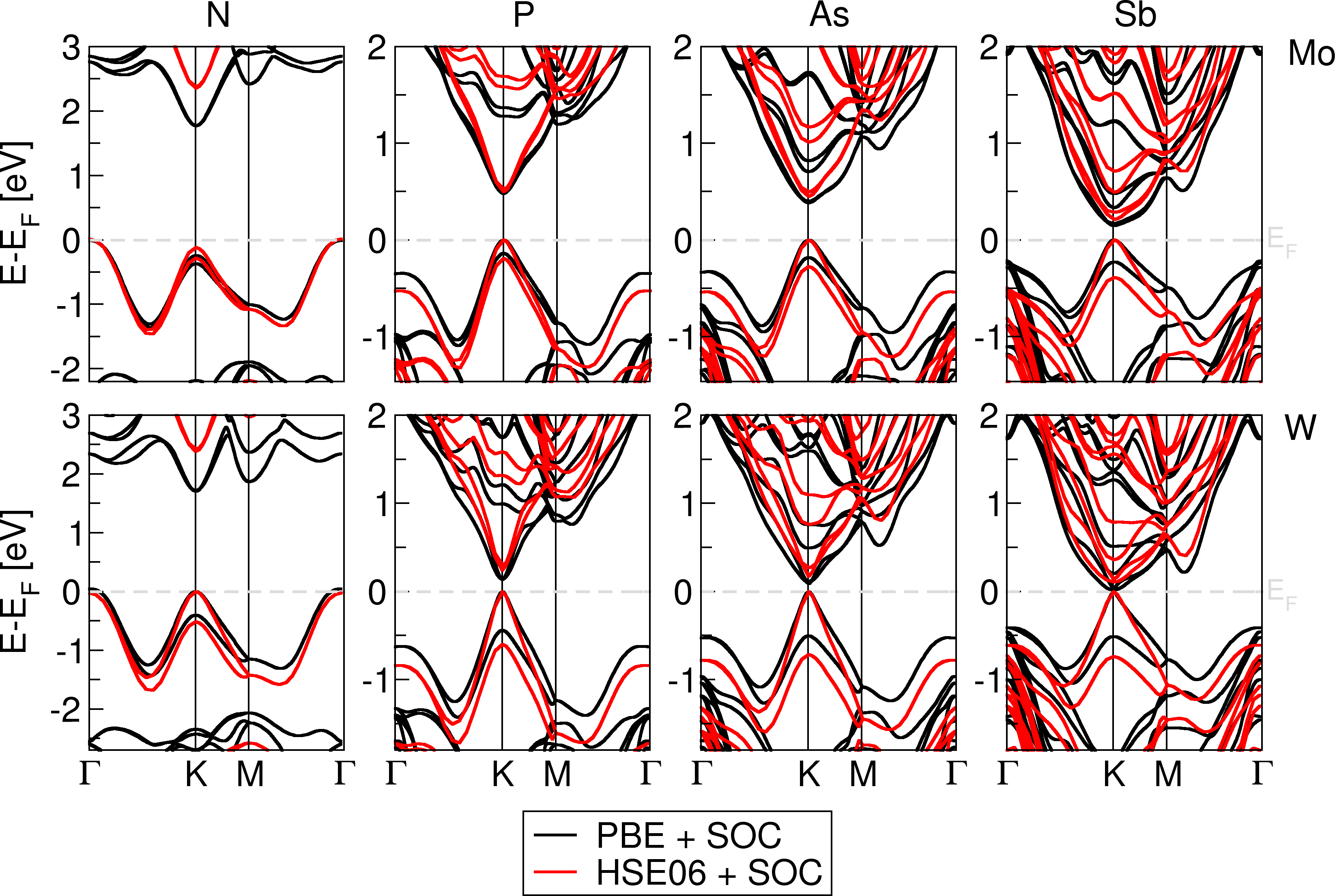}
 \caption{\label{fig:BS_PBE_HSE} Electronic band structures of all compounds calculated at the PBE (black) and HSE06 (red) level of theory including SOC.}
\end{figure}

  \begin{figure}[ht!]\centering
\includegraphics[width=0.6\textwidth]{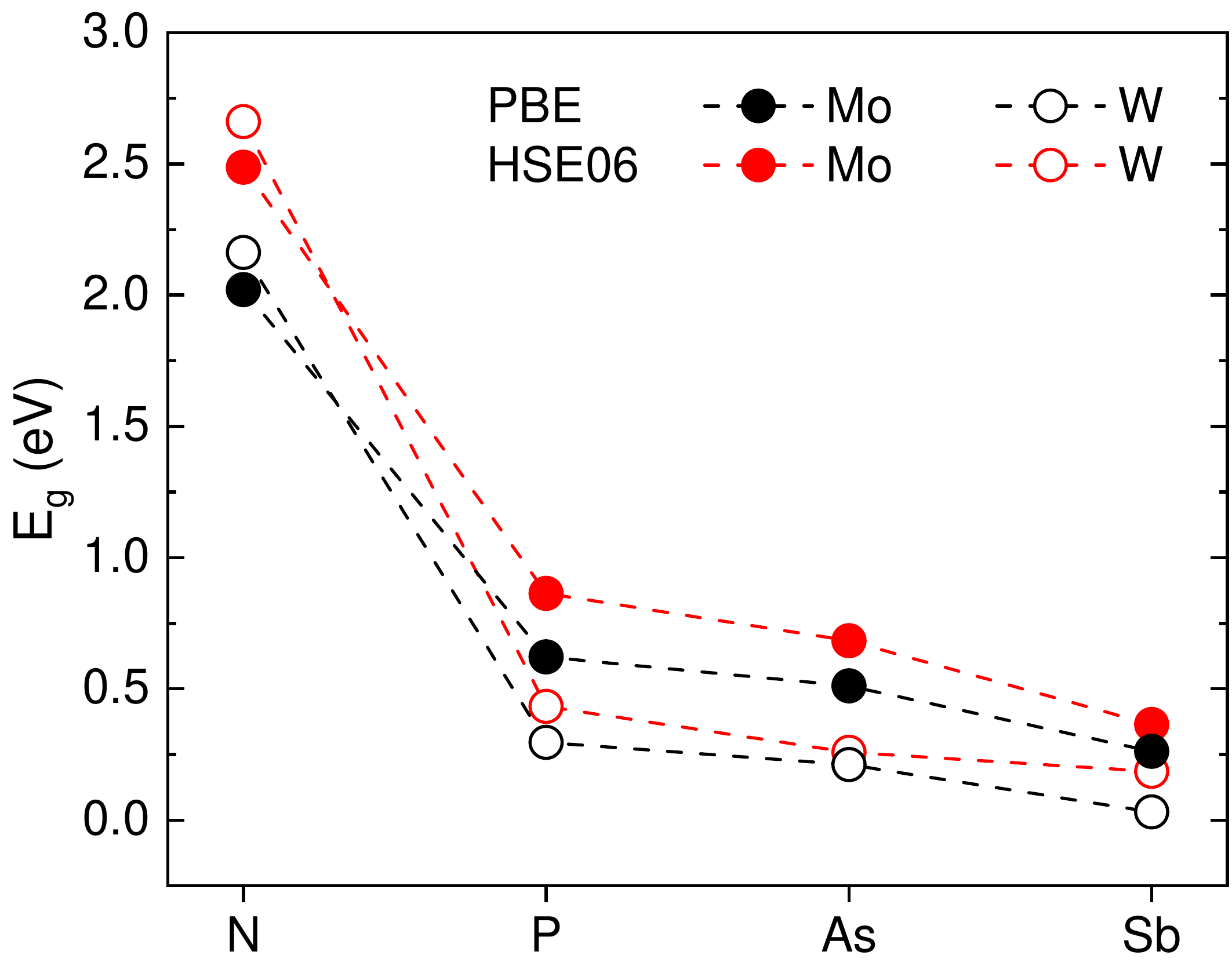}
 \caption{\label{fig:gap_trends} Energies of the direct band gaps at $K$ for Mo- (solid circles) and W-based (hollow circles) compounds calculated at the PBE (black) and HSE06 (red) level of theory with SOC.}
\end{figure}

  \begin{figure}[ht!]\centering
\includegraphics[width=\textwidth]{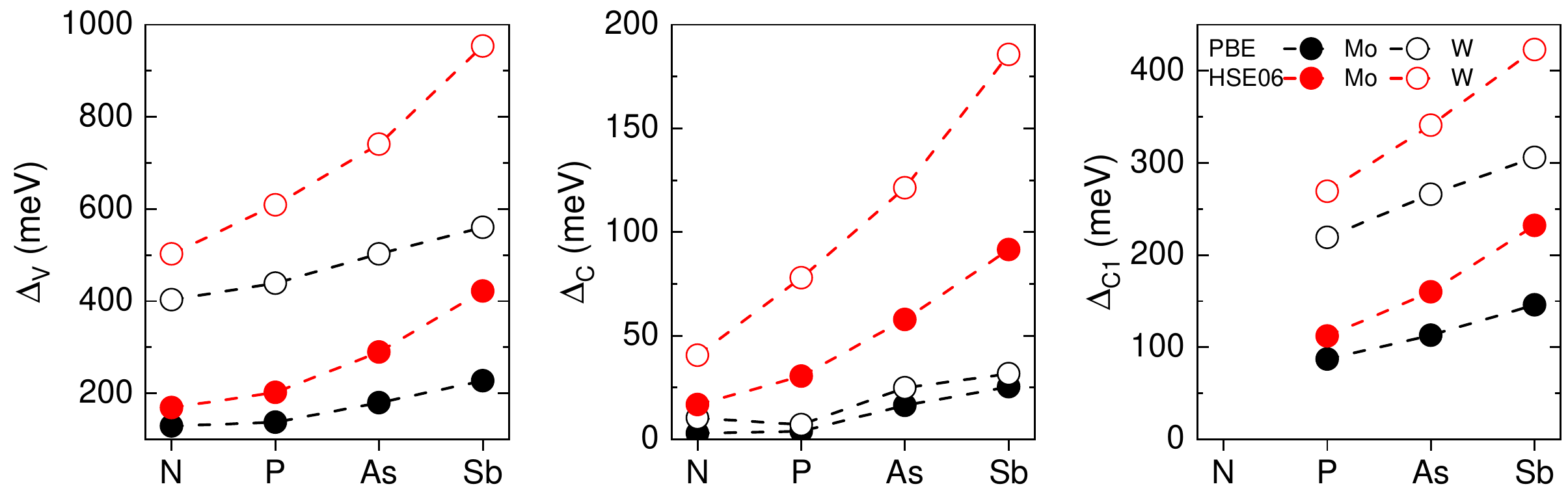}
 \caption{\label{fig:splitting_trends} Spin-splittings of $v$, $c$ and $c1$ bands at $K$ for Mo- (solid circles) and W-based (hollow circles) compounds calculated at the PBE (black) and HSE06 (red) level of theory with SOC.}
\end{figure}

  \begin{figure}[ht!]\centering
 \includegraphics[width=0.75\textwidth]{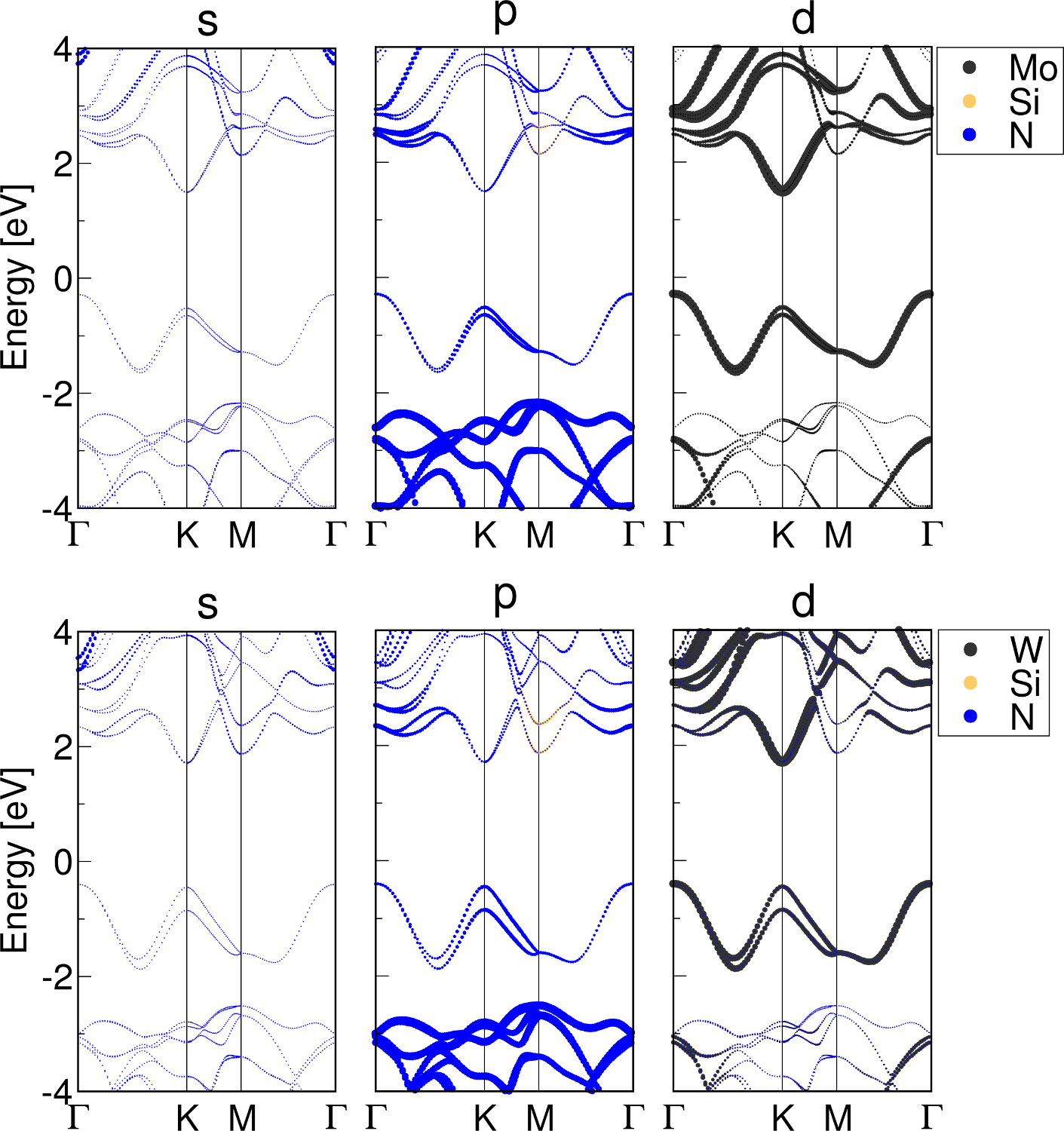}
 \caption{\label{fig:fatbands_MoSi2N4_WSi2N4} Electronic band structures of MoSi$_2$N$_4$ and WSi$_2$N$_4$ with contribution of atomic orbitals.}
\end{figure}


 \begin{figure}[ht!]\centering
 \includegraphics[width=0.75\textwidth]{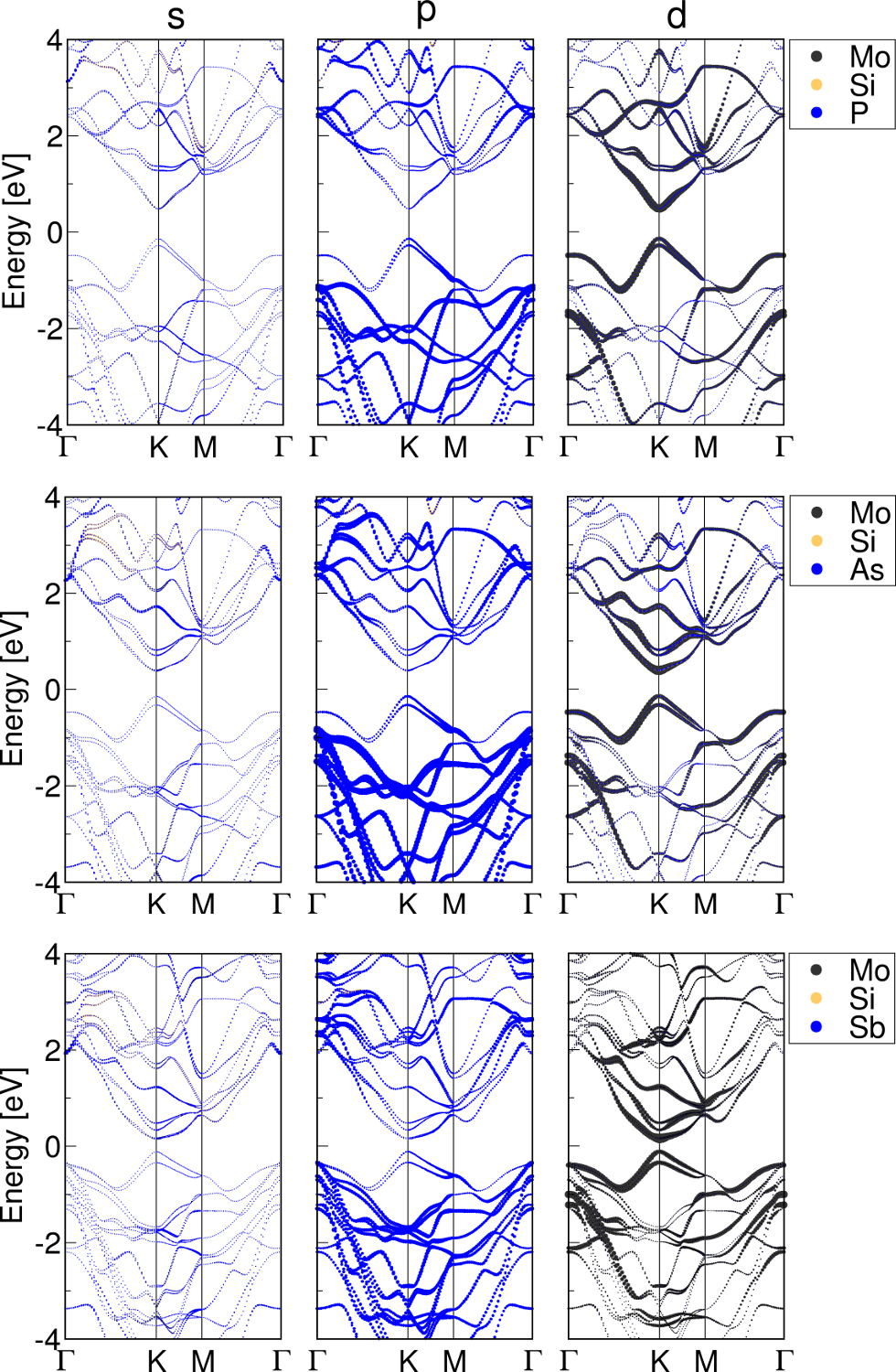}
 \caption{\label{fig:fatbands_Mo} Electronic band structures of MoSi$_2Z_4$, (Z = P, As, Sb) with contribution of atomic orbitals.}
\end{figure}

 \begin{figure}[ht!]\centering
 \includegraphics[width=0.75\textwidth]{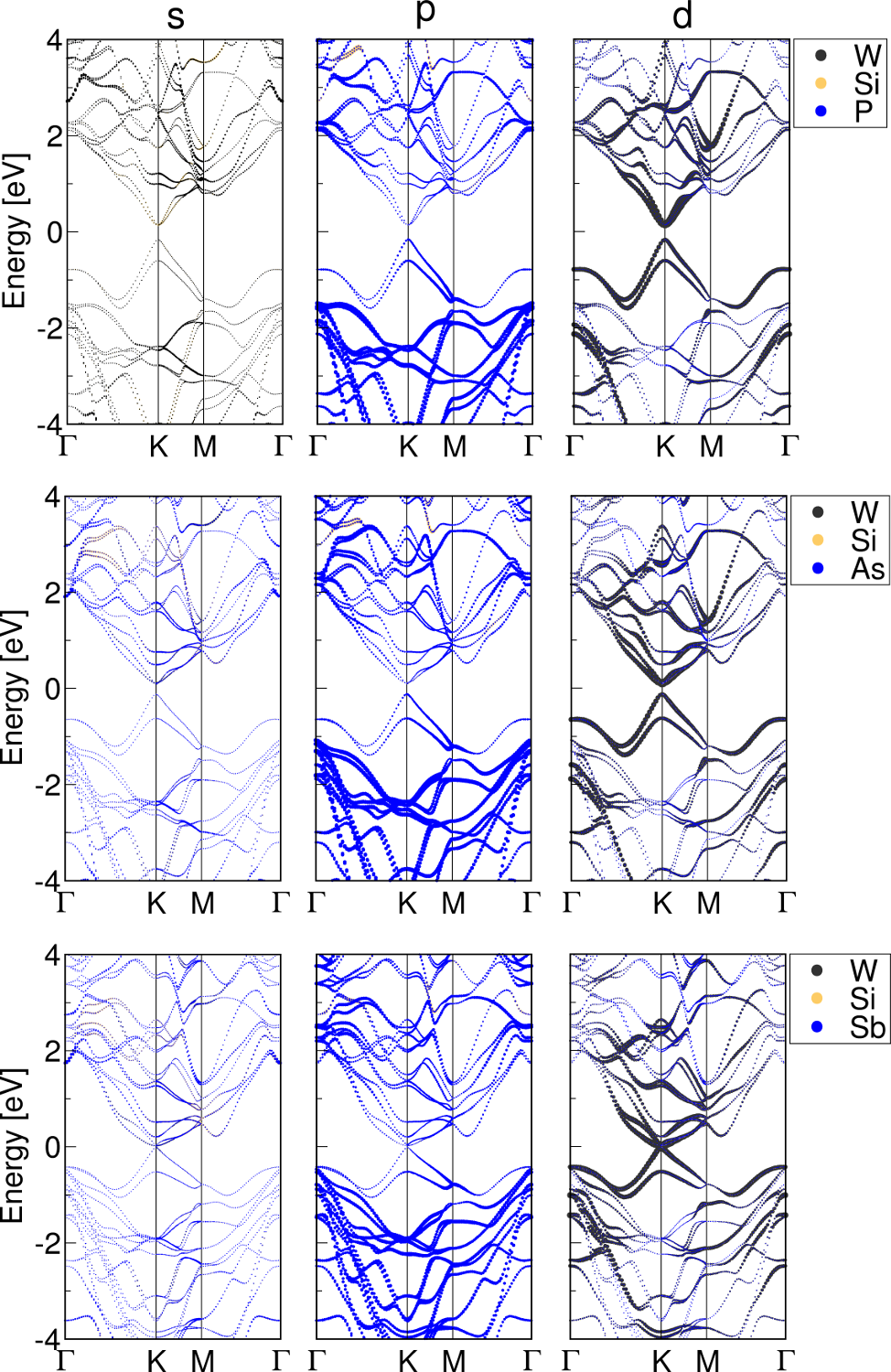}
 \caption{\label{fig:fatbands_W} Electronic band structures of WSi$_2Z_4$, (Z = P, As, Sb) with contribution of atomic orbitals.}
\end{figure}


 \begin{figure}[ht!]\centering
 \includegraphics[width=\textwidth]{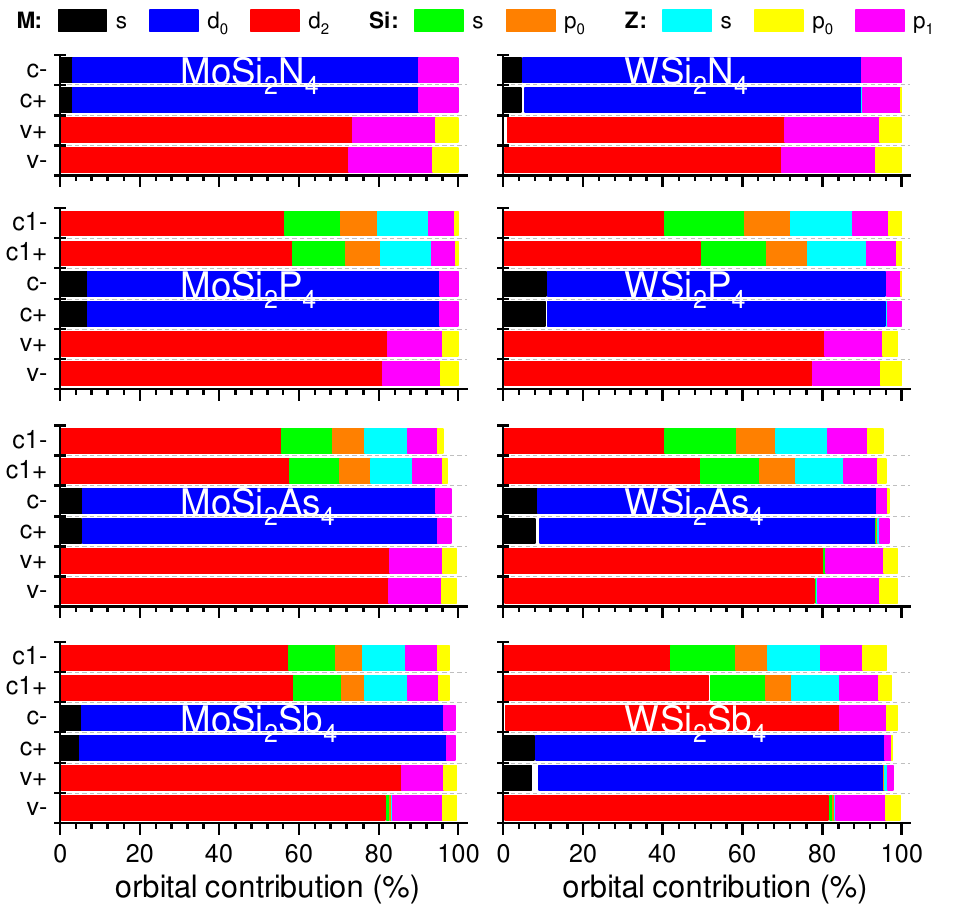}
 \caption{\label{fig:orbitals_K} Orbital compositions of two uppermost valence bands $v\pm$ and four lowermost conduction bands $c\pm$ and $c1\pm$ at the $K$ point in 1L $M$Si$_{\textrm{2}}Z_{\textrm{4}}$ compounds. For clarity, the orbitals with minor contributions are not plotted. }
\end{figure}

  \begin{figure}[ht!]\centering
\includegraphics[width=\textwidth]{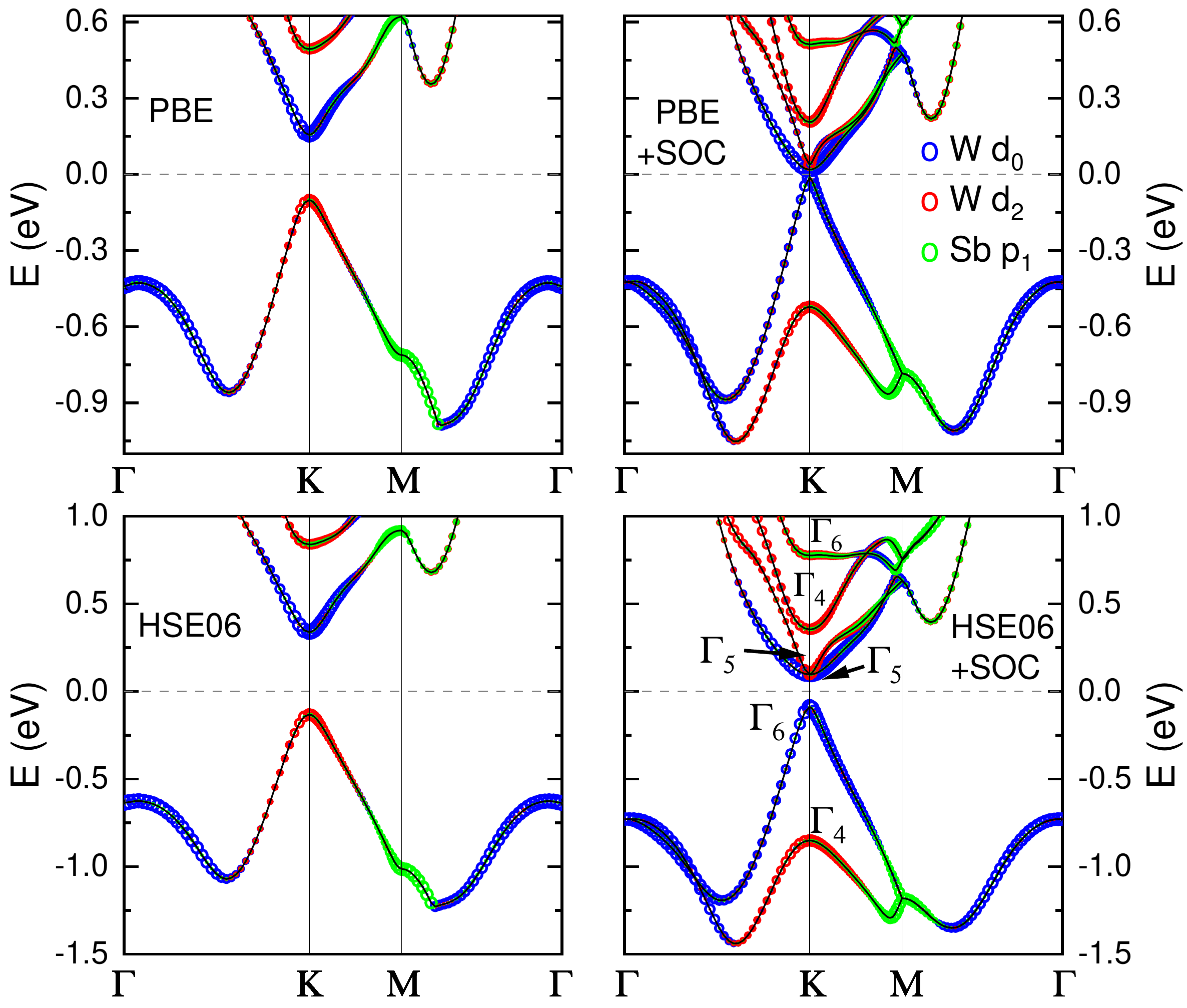}
 \caption{\label{fig:WSi2Sb4} Band structures of WSi$_2$Sb$_4$ with atomic orbital contributions from PBE and HSE06 calculations without and with SOC, which show a SOC-induced band inversion.}
\end{figure}


 \begin{figure}[ht!]\centering
 \includegraphics[width=0.75\textwidth]{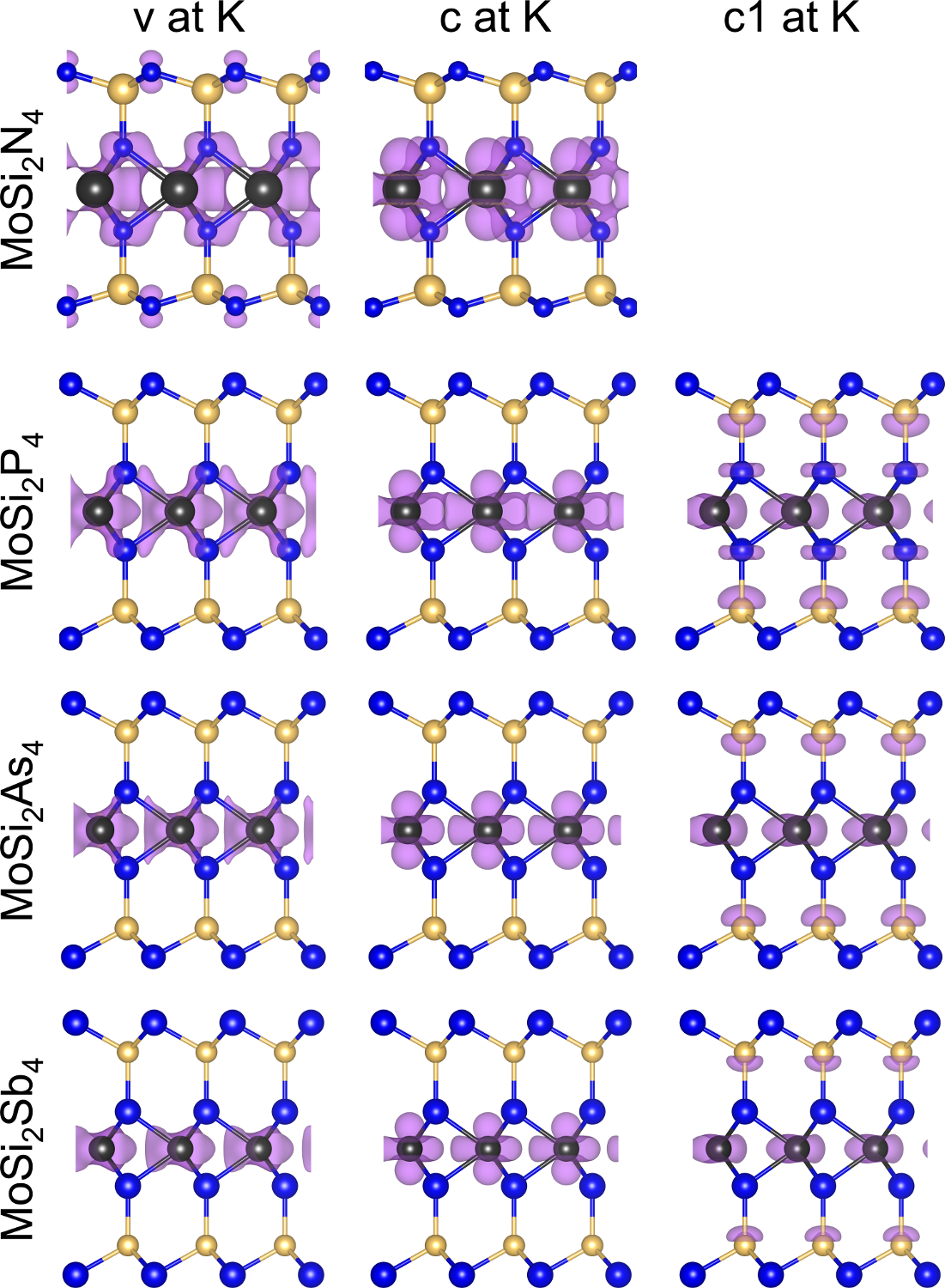}
 \caption{\label{fig:PCD_Mo} Electron and hole states of all 1L MoSi$_2Z_4$ materials at the $K$ point and their wave function extents in real space - isosurfaces of the partial charge density with isovalue of 0.007~e~\AA$^{-3}$. $v$ - valence band maximum, $c$ - conduction band minimum, $c1$ - higher energy conduction band at $K$.}
\end{figure}

 \begin{figure}[ht!]\centering
 \includegraphics[width=0.75\textwidth]{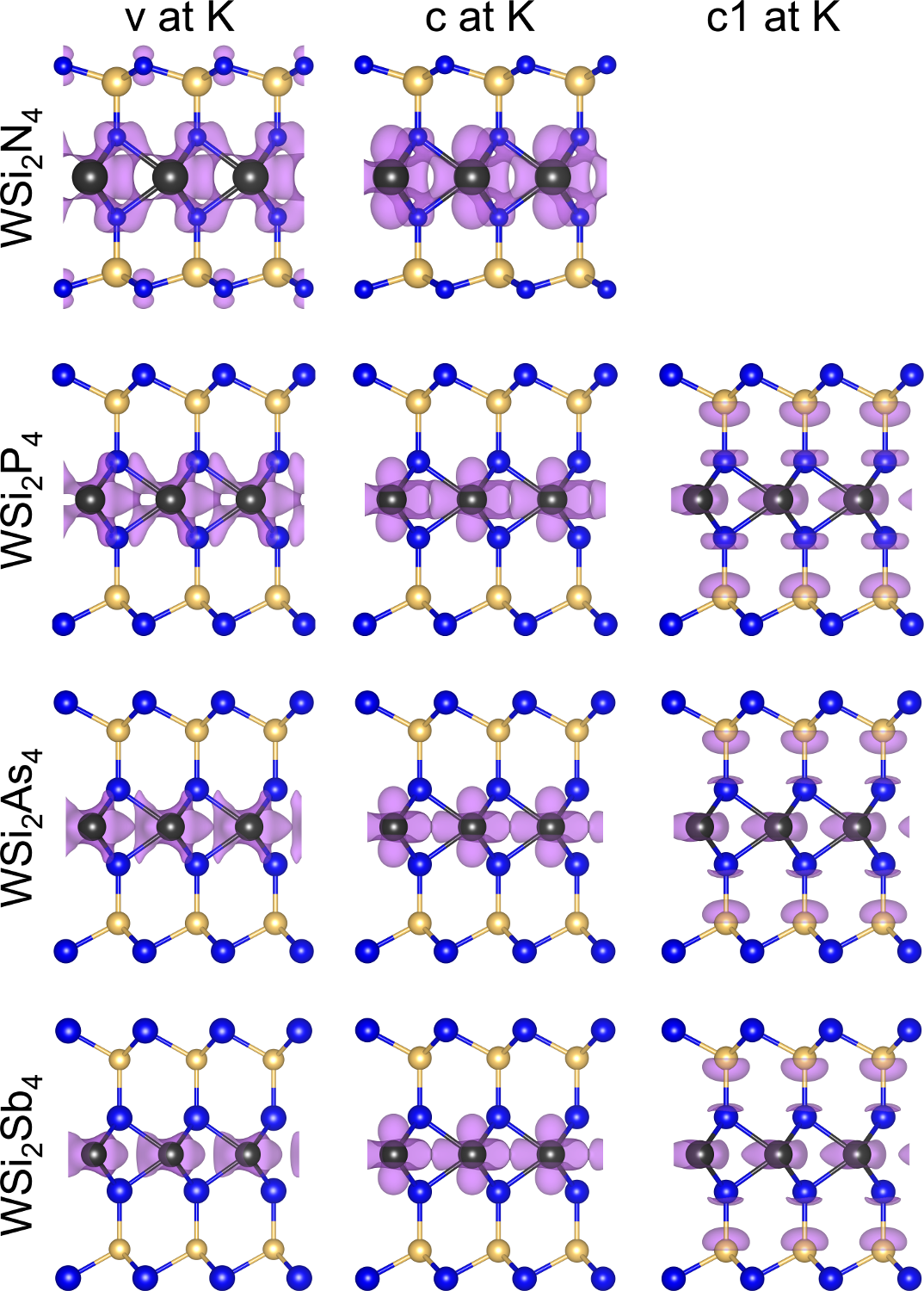}
 \caption{\label{fig:PCD_W} Electron and hole states of all 1L WSi$_2Z_4$ materials at the $K$ point and their wave function extents in real space - isosurfaces of the partial charge density with isovalue of 0.007~e~\AA$^{-3}$. $v$ - valence band maximum, $c$ - conduction band minimum, $c1$ - higher energy conduction band at $K$.}
\end{figure}


 \begin{figure}[ht!]\centering
 \includegraphics[width=1\textwidth]{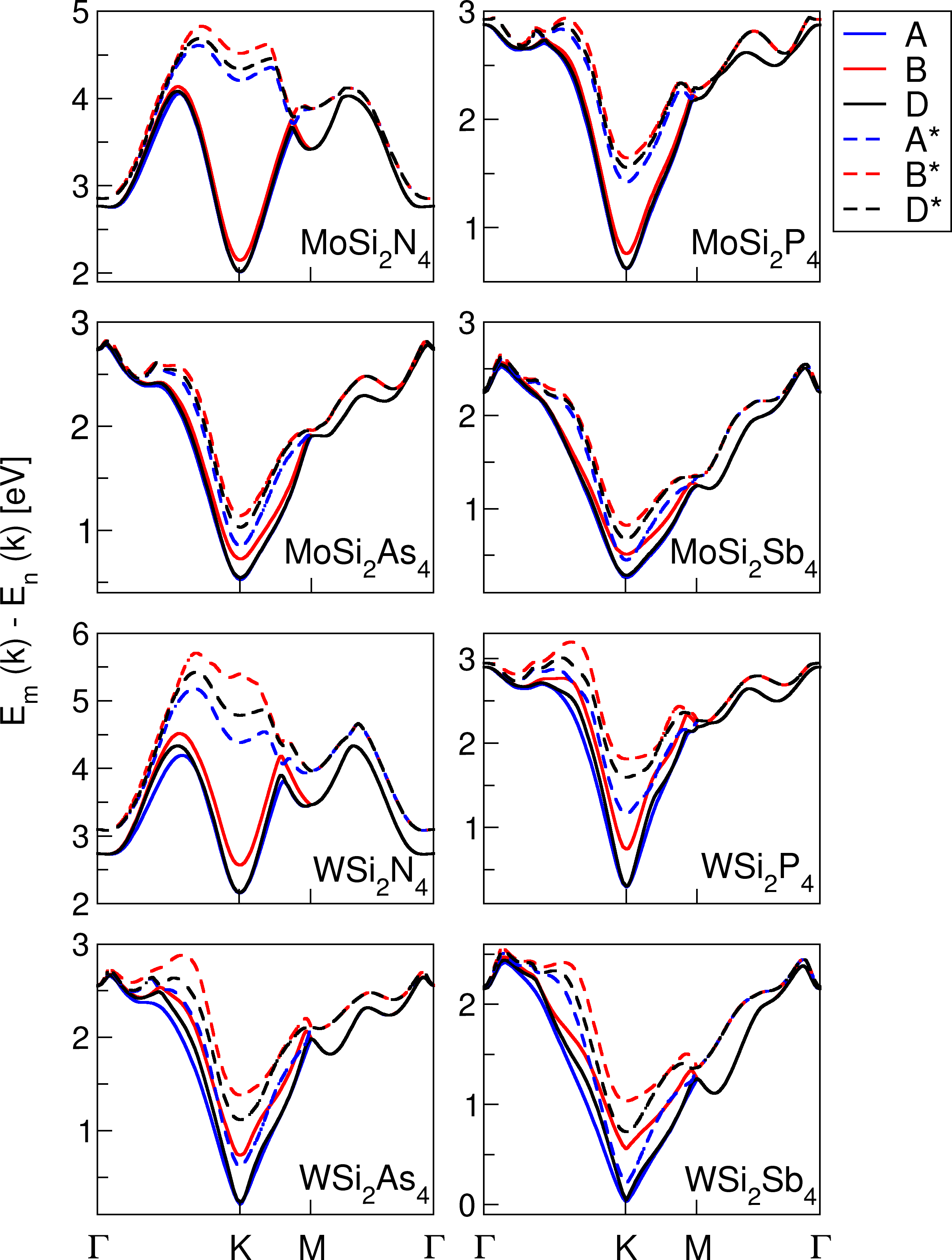}
 \caption{\label{fig:Ediff_PBE} Energy difference $E_m(k) - E_n(k)$ plots for all the compounds calculated at the PBE level of theory with SOC.}
\end{figure}

 \begin{figure}[ht!]\centering
 \includegraphics[width=1\textwidth]{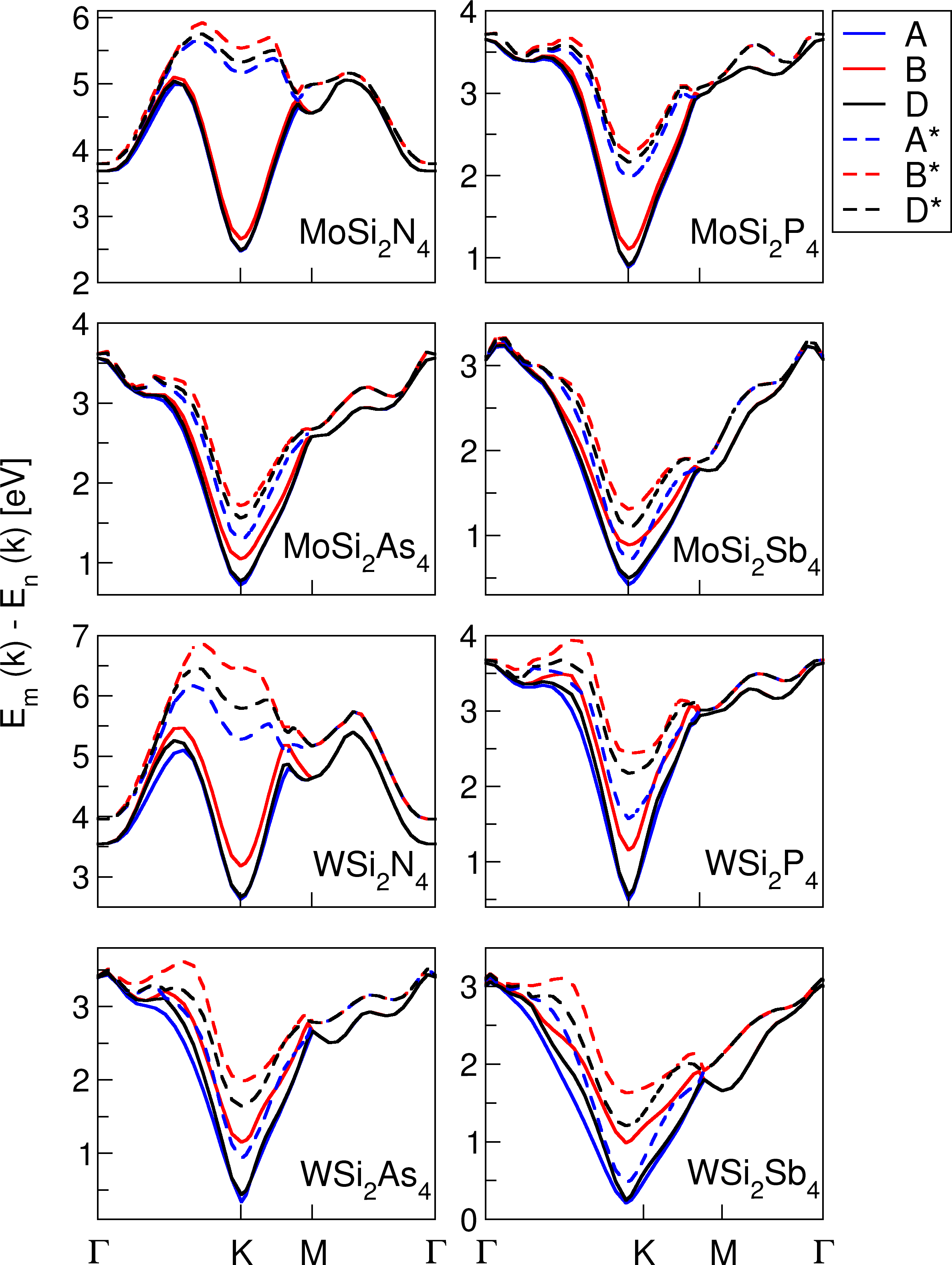}
 \caption{\label{fig:Ediff_HSE} Energy difference $E_m(k) - E_n(k)$ plots for all the compounds calculated at the HSE06 level of theory with SOC.}
\end{figure}


 \begin{figure}[ht!]\centering
 \includegraphics[width=\textwidth]{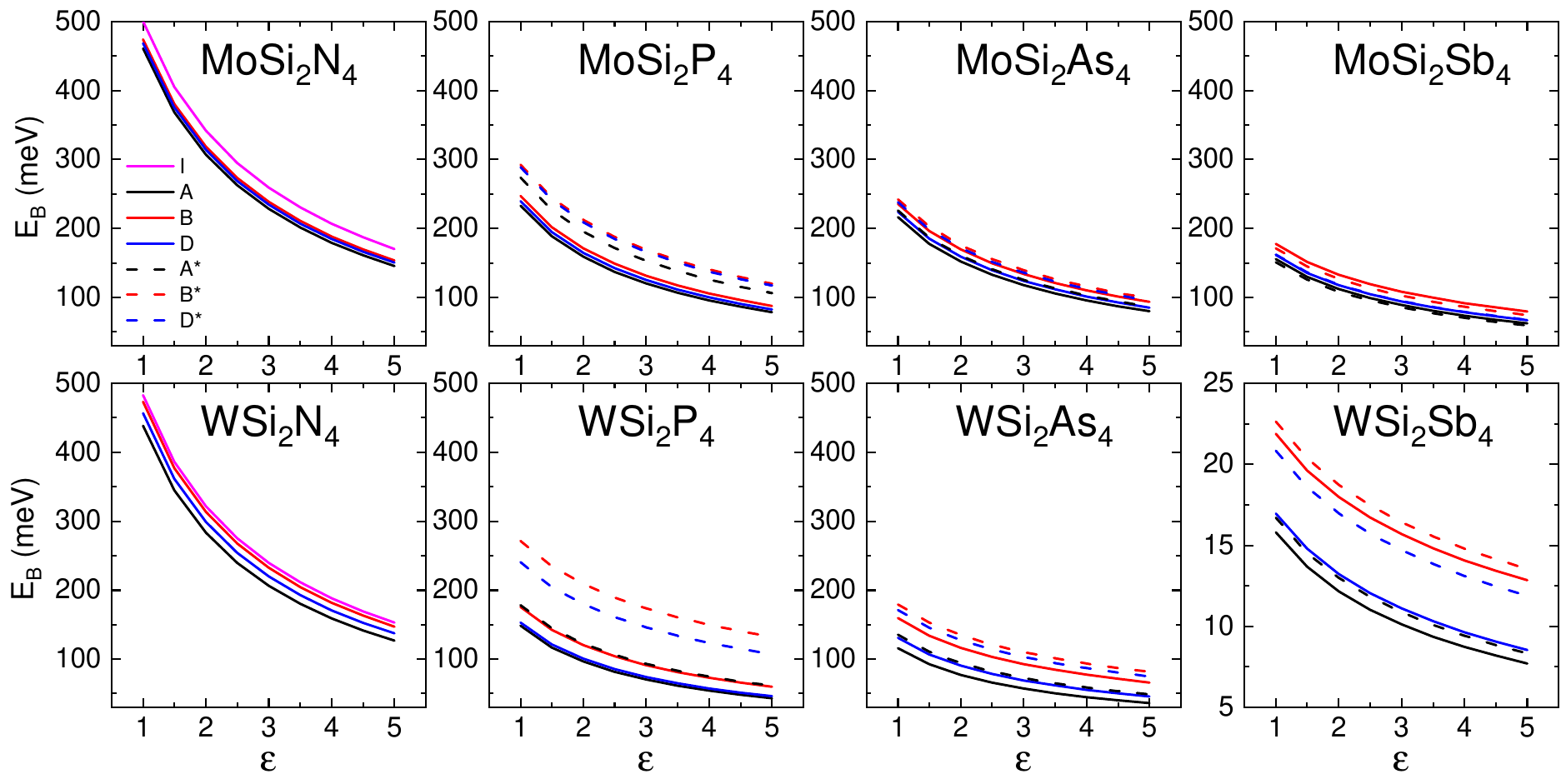}
 \caption{\label{fig:Eb_all} Exciton binding energies, $E_B$, of I, A, B, D, A*, B* and D* excitons in 1L $M$Si$_{\textrm{2}}Z_{\textrm{4}}$ compounds as function of the effective dielectric constant $\varepsilon$.}
\end{figure}

\clearpage
\subsection{Supplementary Tables}

 \begin{table}[ht!]\centering
 \caption{\label{tab:m_rho}  Effective masses of bands $m_n$, $n=\Gamma, v\pm$, $c\pm$, $c1\pm$ valleys and screening lengths $r_0$ calculated from PBE level of theory. For MoSi$_2$N$_4$ and WSi$_2$N$_4$, $m_{v,\Gamma}$ values are equal to -1.66 and -1.55, respectively.}
\begin{tabular}{lcccccccc}
\hline
     & $v_\Gamma$ & $v-$    & $v+$    & $c+$    & $c-$    & $c1+$   & $c1-$ & $r_0$ \\
    \hline
    MoSi$_2$N$_4$  & -1.66 & -0.61 & -0.55 & 0.40  & 0.45  &       &       & 51.6 \\
    MoSi$_2$P$_4$  &      & -0.39 & -0.30 & 0.28  & 0.35  & 8.55  & -32.16 & 109.2 \\
    MoSi$_2$As$_4$ &      & -0.50 & -0.34 & 0.42  & 0.62  & 0.69  & 0.86  & 130.5 \\
    MoSi$_2$Sb$_4$ &      & -0.60 & -0.30 & 0.54  & 1.03  & 0.39  & 0.65  & 200.3 \\
    \hline
    WSi$_2$N$_4$  & -1.55 & -0.56 & -0.41 & 0.29  & 0.38  &       &       & 49.4 \\
    WSi$_2$P$_4$  &       & -0.35 & -0.11 & 0.13  & 0.17  & -1.21 & -0.55 & 146.8 \\
    WSi$_2$As$_4$ &       & -0.48 & -0.12 & 0.11  & 0.43  & 1.01  & 2.90  & 202.3 \\
    WSi$_2$Sb$_4$ &       & -0.65 & -0.07 & 0.71  & 0.08  & 0.39  & 1.30  & 2280.9 \\
    \hline
    \end{tabular}%
\end{table}

 \begin{table}[ht!]\centering
 \caption{\label{tab:MoSi2N4}   1L MoSi$_2$N$_4$: energies ($E$) calculated at the PBE/HSE06 level of theory, exciton binding energies ($E_B$) from BSE, dipole strengths ($I$), polarizations, exciton $g$-factors, diamagnetic coefficients ($\alpha$), expectation values of squared exciton radii ($\braket{r^2}$), and effective masses of squared excitons ($m_r$) for A, B, D and I transitions.}
    \begin{tabular}{ccccccccc}
    \hline
     transition & $E$ [eV] & $E_B$ [meV] & $I$ [eV$^2$\AA$^2$] & polariz. & $g$ & $\alpha$ [meV T$^{-2}$] & $\sqrt{\braket{r^2}}$ [\AA] & $m_r$ [m$_0$] \\
    \hline
    A     & 2.021/2.487 & 461   & 37.8  & $\sigma+$ & -2.84/-2.82  & 0.15  & 12.4 & 0.22 \\
    B     & 2.153/2.673 & 474   & 36.6  & $\sigma+$ & -2.77/-2.76  & 0.12  & 11.7 & 0.26 \\
    D     & 2.024/2.504 & 469   & 0.003 & z      & -7.26/-7.12  & 0.13  & 12.0 & 0.25 \\
    I     & 1.787/2.350 & 500   &       &        & 5.18/4.26  &       &       &  \\
    \hline
    \end{tabular}%
\end{table}

 \begin{table}[ht!]\centering
 \caption{\label{tab:MoSi2P4}   1L MoSi$_2$P$_4$: energies ($E$) calculated at the PBE/HSE06 level of theory, exciton binding energies ($E_B$) from BSE, dipole strengths ($I$), polarizations, exciton $g$-factors, diamagnetic coefficients ($\alpha$), expectation values of squared exciton radii ($\braket{r^2}$) and effective masses of squared excitons ($m_r$) for A, B, D, A*, B* and D* transitions.}
    \begin{tabular}{ccccccccc}
    \hline
     transition & $E$ [eV] & $E_B$ [meV] & $I$ [eV$^2$\AA$^2$] & polariz. & $g$ & $\alpha$ [meV T$^{-2}$] & $\sqrt{\braket{r^2}}$ [\AA] & $m_r$ [m$_0$]  \\
    \hline
    A     & 0.621/0.864 & 233   & 16.4  & $\sigma+$ & -2.78/-2.82  & 0.80  & 22.5   & 0.14 \\
    B     & 0.762/1.060 & 247   & 15.2  & $\sigma+$ & -2.68/-2.73  & 0.45  & 19.8   & 0.19 \\
    D     & 0.625/0.894 & 240   & 0.001 & $z$       & -8.42/-7.79  & 0.62  & 21.2   & 0.16 \\
    A*    & 1.421/1.954 & 274   & 1.30  & $\sigma-$ & 16.30/14.20  & 0.18  & 15.6   & 0.29 \\
    B*    & 1.645/2.268 & 292   & 1.56  & $\sigma-$ & 14.82/13.40  & 0.10  & 13.3   & 0.40 \\
    D*    & 1.558/2.156 & 289   & 0.017 & $z$       & 10.57/9.14   & 0.11  & 13.7   & 0.38 \\
    \hline
    \end{tabular}%
\end{table}

 \begin{table}[ht!]\centering
 \caption{\label{tab:MoSi2As4}   1L MoSi$_2$As$_4$: energies ($E$) calculated at the PBE/HSE06 level of theory, exciton binding energies ($E_B$) from BSE, dipole strengths ($I$), polarizations, exciton $g$-factors, diamagnetic coefficients ($\alpha$), expectation values of squared exciton radii ($\braket{r^2}$) and effective masses of squared excitons ($m_r$) for A, B, D, A*, B* and D* transitions.}
    \begin{tabular}{ccccccccc}
    \hline
    transition & $E$ [eV] & $E_B$ [meV] & $I$ [eV$^2$\AA$^2$] & polariz. & $g$ & $\alpha$ [meV T$^{-2}$] & $\sqrt{\braket{r^2}}$ [\AA] & $m_r$ [m$_0$]  \\
    \hline
    A     & 0.512/0.684 & 216   & 11.5  & $\sigma+$ & -2.12/-2.17  & 0.52  & 21.3  & 0.19 \\
    B     & 0.708/1.031 & 236   & 11.1  & $\sigma+$ & -2.34/-2.41  & 0.24  & 17.4   & 0.28 \\
    D     & 0.528/0.742 & 224   & 0.0005 & $z$      & -8.15/-7.52  & 0.38  & 19.6   & 0.22 \\
    A*    & 0.890/1.316 & 226   & 1.12  & $\sigma-$ & 13.88/12.24  & 0.35  & 19.2   & 0.23 \\
    B*    & 1.183/1.765 & 242   & 1.14  & $\sigma-$ & 12.31/11.37  & 0.18  & 16.2   & 0.32 \\
    D*    & 1.070/1.605 & 238   & 0.022 & $z$       & 8.11/7.14    & 0.22  & 17.0   & 0.29 \\
    \hline
    \end{tabular}%
\end{table}

 \begin{table}[ht!]\centering
 \caption{\label{tab:MoSi2Sb4}   1L MoSi$_2$Sb$_4$: energies ($E$) calculated at the PBE/HSE06 level of theory, exciton binding energies ($E_B$) from BSE, dipole strengths ($I$), polarizations, exciton $g$-factors, diamagnetic coefficients ($\alpha$), expectation values of squared exciton radii ($\braket{r^2}$) and effective masses of squared excitons ($m_r$) for A, B, D, A*, B* and D* transitions.}
    \begin{tabular}{ccccccccc}
    \hline
     transition & $E$ [eV] & $E_B$ [meV] & $I$ [eV$^2$\AA$^2$] & polariz. & $g$ & $\alpha$ [meV T$^{-2}$] & $\sqrt{\braket{r^2}}$ [\AA] & $m_r$ [m$_0$]  \\
    \hline
    A     & 0.262/0.364 & 155   & 6.3   & $\sigma+$ & -0.71/-0.86  & 0.763 & 25.7  & 0.19 \\
    B     & 0.515/0.877 & 178   & 6.47  & $\sigma+$ & -2.20/-2.24  & 0.189 & 18.1  & 0.38 \\
    D     & 0.288/0.455 & 161   & 0.0002 & $z$      & -8.85/-7.81  & 0.519 & 23.3   & 0.23 \\
    A*    & 0.449/0.712 & 151   & 1.49  & $\sigma-$ & 13.65/11.80  & 0.978 & 27.5   & 0.17 \\
    B*    & 0.822/1.366 & 171   & 1.13  & $\sigma-$ & 11.50/10.87  & 0.284 & 20.0   & 0.31 \\
    D*    & 0.676/1.134 & 162   & 0.03  & $z$       & 7.01/6.24    & 0.489 & 23.1   & 0.24 \\
\hline
    \end{tabular}%
\end{table}

 \begin{table}[ht!]\centering
 \caption{\label{tab:WSi2N4}   1L WSi$_2$N$_4$: energies ($E$) calculated at the PBE/HSE06 level of theory, exciton binding energies ($E_B$) from BSE, dipole strengths ($I$), polarizations, exciton $g$-factors, diamagnetic coefficients ($\alpha$), expectation values of squared exciton radii ($\braket{r^2}$) and effective masses of squared excitons ($m_r$) for A, B, D and I transitions.}
    \begin{tabular}{ccccccccc}
    \hline
     transition & $E$ [eV] & $E_B$ [meV] & $I$ [eV$^2$\AA$^2$] & polariz. & $g$ & $\alpha$ [meV T$^{-2}$] & $\sqrt{\braket{r^2}}$ [\AA] & $m_r$ [m$_0$]  \\
    \hline
    A     & 2.162/2.660 & 438   & 52.7  & $\sigma+$ & -2.31/-2.29  & 0.27  & 14.4   & 0.17 \\
    B     & 2.575/3.203 & 473   & 48.5  & $\sigma+$ & -2.28/-2.29  & 0.14  & 12.3   & 0.23 \\
    D     & 2.172/2.701 & 457   & 0.034 & $z$       & -7.72/-7.33  & 0.19  & 13.3   & 0.20 \\
    I     & 2.108/2.672 & 483   &       &           & 6.94/5.78  &       &       &  \\
    \hline
    \end{tabular}%
\end{table}

 \begin{table}[ht!]\centering
 \caption{\label{tab:WSi2P4}   1L WSi$_2$P$_4$: energies ($E$) calculated at the PBE/HSE06 level of theory, exciton binding energies ($E_B$) from BSE, dipole strengths ($I$), polarizations, exciton $g$-factors, diamagnetic coefficients ($\alpha$), expectation values of squared exciton radii ($\braket{r^2}$) and effective masses of squared excitons ($m_r$) for A, B, D, A*, B* and D* transitions.}
    \begin{tabular}{ccccccccc}
    \hline
     transition & $E$ [eV] & $E_B$ [meV] & $I$ [eV$^2$\AA$^2$] & polariz. & $g$ & $\alpha$ [meV T$^{-2}$] & $\sqrt{\braket{r^2}}$ [\AA] & $m_r$ [m$_0$]  \\
    \hline
    A     & 0.297/0.433 & 148   & 21.5  & $\sigma+$ & -3.59/-3.60  & 6.45  & 42.0  & 0.06 \\
    B     & 0.742/1.120 & 176   & 16.7  & $\sigma+$ & -3.31/-3.34  & 1.71  & 29.3   & 0.11 \\
    D     & 0.304/0.511 & 153   & 0.0013 & $z$      & -20.45/-15.40 & 4.86 & 39.3  & 0.07 \\
    A*    & 1.161/1.599 & 178   & 0.458 & $\sigma-$ & 30.66/24.57  & 1.37  & 28.4   & 0.13 \\
    B*    & 1.819/2.476 & 271   & 1.03  & $\sigma-$ & 19.45/18.44  & 0.02  & 9.7    & 0.95 \\
    D*    & 1.599/2.208 & 241   & 0.121 & $z$       & 13.52/12.50  & 0.08  & 13.6   & 0.48 \\
    \hline
    \end{tabular}%
\end{table}

 \begin{table}[ht!]\centering
 \caption{\label{tab:WSi2As4}   1L WSi$_2$As$_4$: energies ($E$) calculated at the PBE/HSE06 level of theory, exciton binding energies ($E_B$) from BSE, dipole strengths ($I$), polarizations, exciton $g$-factors, diamagnetic coefficients ($\alpha$), expectation values of squared exciton radii ($\braket{r^2}$) and effective masses of squared excitons ($m_r$) for A, B, D, A*, B* and D* transitions.}
    \begin{tabular}{ccccccccc}
    \hline
     transition & $E$ [eV] & $E_B$ [meV] & $I$ [eV$^2$\AA$^2$] & polariz. & $g$ & $\alpha$ [meV T$^{-2}$] & $\sqrt{\braket{r^2}}$ [\AA] & $m_r$ [m$_0$]  \\
    \hline
    A     & 0.211/0.258 & 116   & 15.3  & $\sigma+$ & -3.01/-2.99  & 13.37 & 60.4 & 0.06 \\
    B     & 0.739/1.120 & 159   & 12.1  & $\sigma+$ & -3.05/-3.07  & 0.54  & 23.7   & 0.23 \\
    D     & 0.236/0.380 & 131   & 0.0009 & $z$     & -21.85/-18.70  & 3.50 & 37.9  & 0.09 \\
    A*    & 0.612/0.929 & 135   & 0.464 & $\sigma-$ & 28.06/24.60  & 2.73  & 35.2  & 0.10 \\
    B*    & 1.380/2.010 & 179   & 0.836 & $\sigma-$ & 14.24/13.94  & 0.16  & 17.4   & 0.41 \\
    D*    & 1.115/1.669 & 171   & 0.115 & $z$     & 9.46/9.15  & 0.27  & 19.7   & 0.32 \\
    \hline
    \end{tabular}%
\end{table}

 \begin{table}[ht!]\centering
 \caption{\label{tab:WSi2Sb4}   1L WSi$_2$Sb$_4$: energies ($E$) calculated at the PBE/HSE06 level of theory, exciton binding energies ($E_B$) from BSE, dipole strengths ($I$), polarizations, exciton $g$-factors, diamagnetic coefficients ($\alpha$), expectation values of squared exciton radii ($\braket{r^2}$) and effective masses of squared excitons ($m_r$) for A, B, D, A*, B* and D* transitions.}
    \begin{tabular}{ccccccccc}
    \hline
     transition & $E$ [eV] & $E_B$ [meV] & $I$ [eV$^2$\AA$^2$] & polariz. & $g$ & $\alpha$ [meV T$^{-2}$] & $\sqrt{\braket{r^2}}$ [\AA] & $m_r$ [m$_0$]  \\
    \hline
    A     & 0.051/0.186 & 16    & 7.8   & $\sigma+$ & 3.39/3.41  & 84.10 & 123.7 & 0.04 \\
    B     & 0.541/0.953 & 22    & 6.11  & $\sigma+$ & -39.54/-39.59  & 3.22  & 70.5  & 0.34 \\
    D     & 0.516/0.851 & 17    & 0.069 & $z$     & 16.52/15.70  & 42.96 & 116.9 & 0.07 \\
    A*    & 0.210/0.429 & 17    & 1.88  & $\sigma-$ & -43.99/-43.35  & 51.60 & 118.7 & 0.06 \\
    B*    & 1.025/1.619 & 23    & 0.964 & $\sigma-$ & 6.67/6.54  & 2.13  & 64.6  & 0.43 \\
    D*    & 0.719/1.197 & 21    & 0.115 & $z$     & 34.14/34.19  & 5.76  & 81.0  & 0.25 \\
\hline
    \end{tabular}%
\end{table}

 \begin{table}[ht!]\centering
 \caption{\label{tab:L} Orbital angular momenta $L_{n,+K}$ of bands $n=v\pm$, $c\pm$, $c1\pm$ at $K+$ point calculated from Equation~2 in main text using PBE/HSE06 band gaps. For MoSi$_2$N$_4$ and WSi$_2$N$_4$, $L_{v,\Gamma}$ values are equal to 0.009/0.005 and 0.031/0.016, respectively. }
\begin{tabular}{lccccccc}
\hline
     &  $v-$    & $v+$    & $c+$    & $c-$    & $c1+$   & $c1-$ \\
    \hline
    MoSi$_2$N$_4$   & 3.78/3.37 & 4.02/3.55 & 2.60/2.14 & 2.39/1.99 & & \\
    MoSi$_2$P$_4$        & 4.35/3.73 & 5.22/4.25 & 3.83/2.84 & 3.01/2.36 & -2.93/-2.84 & -3.05/-2.97 \\
    MoSi$_2$As$_4$       & 3.87/3.47 & 4.75/4.03 & 3.69/2.94 & 2.68/2.67 & -2.19/-2.10 & -2.29/-2.21 \\
    MoSi$_2$Sb$_4$       & 3.48/3.22 & 4.80/4.01 & 4.45/3.58 & 2.38/2.11 & -2.02/-1.88 & -2.27/-2.21 \\
    \hline
    WSi$_2$N$_4$   & 3.94/3.53 & 4.66/4.05 & 3.50/2.91 & 2.80/2.39 & & \\
    WSi$_2$P$_4$        & 4.86/4.39 & 11.43/8.43 & 9.64/6.63 & 3.21/2.73 & -3.90/-3.85 & -4.87/-4.83 \\
    WSi$_2$As$_4$       & 4.11/3.98 & 11.41/9.71 & 9.88/8.17 & 2.49/2.36 & -2.62/-2.59 & -3.01/-2.99 \\
    WSi$_2$Sb$_4$       & 0.57/0.59 & 3.49/3.17 & 1.80/1.46 & -19.2/-19.2 & -18.5/-18.5 & -2.76/-2.68 \\
    \hline
    \end{tabular}%
\end{table}

\end{document}